\documentclass[a4paper,11pt]{article}

\usepackage{jheppub}
\allowdisplaybreaks
\usepackage{graphicx}
\usepackage{epstopdf}
\usepackage{amsthm}

\newcommand{\be}{\begin{equation}}
\newcommand{\ee}{\end{equation}}
\newcommand{\bea}{\begin{eqnarray}}
\newcommand{\eea}{\end{eqnarray}}
\newcommand{\nn}{\nonumber}
\def\nl{\nonumber \\ &}

\title{Spinning gravitating objects in the\\ effective field theory in the post-Newtonian scheme}

\author[a,b]{Michele Levi}
\author[c,d]{and Jan Steinhoff}

\affiliation[a]{Universit\'e Pierre et Marie Curie-Paris VI, CNRS-UMR 7095, 
Institut d'Astrophysique de Paris,\\ 98 bis Boulevard Arago, 75014 Paris, France} 
\affiliation[b]{Sorbonne Universit\'es, Institut Lagrange de Paris,\\ 
98 bis Boulevard Arago, 75014 Paris, France} 
\affiliation[c]{Max-Planck-Institute for Gravitational Physics 
(Albert-Einstein-Institute),\\ Am M{\"u}hlenberg 1, 14476 Potsdam-Golm, Germany}
\affiliation[d]{Centro Multidisciplinar de Astrofisica, Instituto Superior Tecnico, 
Universidade de Lisboa,\\ Avenida Rovisco Pais 1, 1049-001 Lisboa, Portugal}

\emailAdd{michele.levi@upmc.fr}
\emailAdd{jan.steinhoff@aei.mpg.de}

\abstract{We introduce a formulation for spinning gravitating objects in the effective field theory in the post-Newtonian scheme in the context of the binary inspiral problem. We aim at an effective action, where all field modes below the orbital scale are integrated out. We spell out the relevant degrees of freedom, in particular the rotational ones, and the associated symmetries. Building on these symmetries, we introduce the minimal coupling part of the point particle action in terms of gauge rotational variables, and construct the spin-induced nonminimal couplings, where we obtain the leading order couplings to all orders in spin. We specify the gauge for the rotational variables, where the unphysical degrees of freedom are eliminated already from the Feynman rules, and all the orbital field modes are integrated out. The equations of motion of the spin can be  directly obtained via a proper variation of the action, and Hamiltonians may be straightforwardly derived. We implement this effective field theory for spin to derive all spin dependent potentials up to next-to-leading order to quadratic level in spin, namely up to the third post-Newtonian order for rapidly rotating compact objects. In particular, the proper next-to-leading order spin-squared potential and Hamiltonian for generic compact objects are also derived. For the implementations we use the nonrelativistic gravitational field decomposition, which is found here to eliminate higher-loop Feynman diagrams also in spin dependent sectors, and facilitates derivations. This formulation for spin is thus ideal for treatment of higher order spin dependent sectors.} 

\begin{document}

\maketitle

\flushbottom

\section{Introduction}

The anticipated direct detection of gravitational waves (GWs) may be 
realized soon with the upcoming operation of second-generation ground-based 
interferometers, such as the twin Advanced LIGO \cite{LIGO} detectors in the 
US, Advanced Virgo \cite{Virgo} in Europe, and in a few years also KAGRA 
\cite{Kagra} in Japan. This will open a new era of observational gravitational 
wave astronomy, where also space-based detectors, such as eLISA 
\cite{Lisa, eLisa} are planned to extend the observed frequency range to the 
low frequency band. Binaries of compact objects are the most promising sources 
in the accessible frequency band of such experiments. The post-Newtonian (PN) 
approximation of General Relativity stands out among the various and 
complementary approaches to model these systems, as it enables to treat 
analytically the inspiral phase of their evolution 
\cite{Blanchet:2013haa}. 

The search for GW signals from such sources employs the matched-filtering 
technique, and thus accurate theoretical template waveforms are crucial to 
obtain a successful detection. Even relative high order PN corrections, such 
as the fourth PN (4PN) order, have an impact on the waveform templates for 
the binary inspiral, and further they are required to gain information about 
the inner structure of the components of the binary \cite{Yagi:2013baa}. 
Moreover, astrophysical observations indicate that such black hole 
components have near extreme spin \cite{McClintock:2011zq}. Hence, PN spin 
effects for rapidly rotating compact objects, which first appear at 1.5PN 
order, should be obtained at least to 4PN order, which was recently 
completed in the non-spinning case \cite{Damour:2014jta}.

Several efforts have been made in recent years to push ahead the formulation 
for gravitating spinning objects in the context of the binary inspiral 
problem. An action formalism plays a central role in the various approaches, 
building in particular on the seminal works in \cite{Hanson:1974qy} and 
\cite{Bailey:1975fe} for flat and curved spacetime, respectively, and see also 
section 11 of \cite{Blanchet:2013haa} for a review of spinning compact 
binaries for gravitational radiation. The self-contained Effective Field 
Theory (EFT) approach for the binary inspiral as introduced in 
\cite{Goldberger:2004jt,Goldberger:2007hy} for non-spinning objects seems 
then to provide a solid path to obtain such a formulation. In the EFT formulation 
manifest power counting in the small expansion parameter (here $v\ll c=1$) is achieved 
by performing a decomposition of the gravitational field at the level of the action 
into modes with definite scaling properties, followed by integrating out the off-shell 
modes \cite{Goldberger:2004jt}. 

The EFT approach provides a systematic methodology 
to construct the action to arbitrarily high accuracy, in terms of operators with 
Wilson coefficients ordered by relevance, which is indispensable beyond the point-mass 
approximation. In that respect it should be pointed out, that even just the point-mass approximation, which past work was using, is naturally incorporated already in the EFT framework. It should also be noted that some effective action with derivative expansion to model finite size effects was already discussed in \cite{Damour:1998jk} in the context of alternative theories of gravity. The EFT approach also provides a natural framework to handle the regularization required for higher order corrections in the PN approximation within the standard renormalization scheme. It formulates the perturbative 
calculation efficiently by applying the standard tools from Quantum Field 
Theory, such as Feynman diagrams (a related basic diagrammatic expansion was used already in \cite{Damour:1995kt}). Consequently, the EFT approach then benefits from existent developed Feynman integral calculus at its disposal. For spinning objects such EFT techniques were first used in \cite{Porto:2005ac}, and revised in \cite{Porto:2008tb}, where eventually a Routhian approach from \cite{Yee:1993ya} was adopted.

Our goal in this work is indeed to obtain an EFT formulation for 
gravitating spinning objects for the binary inspiral problem, building on 
\cite{Goldberger:2004jt,Goldberger:2007hy}, and on several observations made in a series of works, mainly \cite{Levi:2008nh,Levi:2010zu}, \cite{Hergt:2011ik} and \cite{Levi:2014sba}. 
The essential obstacle that one has to deal with in extending the 
formulation from a gravitating point-mass to a spinning object in terms of an 
EFT point particle approach is just the intrinsic conflict between the actual 
spinning object, which must be extended for its rotational velocity not to be 
superluminal, and its view as a point particle. The elusive notion of a 
`center', which would serve as a reference point within the object, in 
relativistic physics, similar to the center of mass in Newtonian physics, is 
the origin of ambiguities in the description of relativistic spinning objects. 
Ever since the first treatment in 1959 of the leading order (LO) PN correction, 
which involves spin, in the spin-orbit sector, the essential choice of such a 
center has been a puzzling issue, see Tulczyjew's paper and errata in 
\cite{Tulczyjew:1959}.  

In this work we aim at an effective action, that incorporates the essential requirement, that all field modes below the 
orbital scale are integrated out. We aim to attain accuracy at the 4PN order for rapidly rotating 
compact objects, and indeed the formulation in this paper holds as it stands 
to this high PN order, and it may hold until dissipative effects start to 
play a role as of the 5PN order \cite{Poisson:2004cw}. 
Here, we spell out the relevant degrees of freedom 
(DOFs), in particular the rotational ones, and most importantly the associated 
symmetries. Building on these symmetries, we start with the minimal coupling 
part of the point particle action, stressing the role of the worldline spin 
as a further worldline rotational DOF. We proceed to construct the 
spin-induced nonminimal couplings, where we obtain the LO couplings to all 
orders in spin. We then introduce the gauge freedom of 
the rotational variables into the action, and express it in terms of gauge 
rotational variables. Again, this spin gauge invariance was not addressed 
previously in the action. From introducing this spin gauge freedom we get that 
the minimal coupling part of the spin in the action, would contribute to the 
finite size effects, which is just the manifestation of the aforementioned 
conflict between the actual spinning extended object and its view as a point 
particle. We then fix a canonical gauge for the rotational variables, where 
the unphysical DOFs are eliminated already from the Feynman rules, and all 
the orbital field modes are conveniently integrated out. 

The equations of motion (EOM) of the spin are then directly obtained via a proper 
variation of the action, where they take on a simple form. 
The corresponding Hamiltonians are also straightforwardly obtained from the potentials, derived via this formulation, due to the canonical gauge fixing.
We implement this EFT formulation for spin to derive all spin dependent potentials up to 
next-to-leading order (NLO) to quadratic level in spin, i.e.~up to the 3PN order for 
rapidly rotating compact objects. 
In particular, the proper next-to-leading order spin-squared potential and Hamiltonian 
for generic compact objects are also derived.
For the implementations we use the nonrelativistic gravitational (NRG) field 
decomposition \cite{Kol:2007bc, Kol:2010ze}, which is found here to eliminate 
higher-loop Feynman diagrams also for spin dependent sectors, and facilitates derivations. 
Hence, with the simple EOM of the spin, and the additional advantageous 
usefulness of the Hamiltonian for the straightforward obtainment of 
gauge-invariant quantities, and for implementations within the effective 
one-body formulation \cite{Buonanno:1998gg}, the EFT formulation for spin here 
is ideal for treatment of higher order spin dependent sectors. Indeed, the application of the EFT formulation for spin presented here has led to the completion of the spin dependent conservative sector up to the 4PN order in the recent works \cite{Levi:2014gsa}, \cite{Levi:2015uxa}, and \cite{Levi:2015ixa}, which obtained the LO cubic and quartic in spin, NNLO spin-orbit, and NNLO spin-squared sectors, respectively. 

The outline of the paper is as follows. In section \ref{setup} we present the 
setup and goal of our EFT formulation for gravitating spinning objects, and 
detail the relevant DOFs, and the associated symmetries. In section 
\ref{formulationeftfors} we start by presenting the minimal coupling part of 
the action, and then express it in terms of rotational gauge variables, which 
yields an extra term from minimal coupling. In section \ref{nmcs} we construct 
the spin-induced nonminimal coupling part of the action, where we obtain the 
LO couplings to all orders in spin. In section 
\ref{integrateoutorbital} we fix all ingredients in order to integrate out 
the orbital field modes: we disentangle the tetrad field from the worldline 
tetrad, we fix the gauge of the tetrad field and of the rotational variables, 
and present the resulting Feynman rules. We also discuss how the EOM of the spin 
are then directly obtained after the orbital modes have been integrated out.
In section \ref{implementation} we implement this EFT for spin to derive all 
spin dependent potentials and Hamiltonians up to NLO to quadratic level in spin, 
i.e.~up to the 3PN order for rapidly rotating compact objects. In section 
\ref{theendmyfriend} we summarize our main conclusions.

Throughout this paper we use $c\equiv1$, 
$\eta_{\mu\nu}\equiv \text{Diag}[1,-1,-1,-1]$,
and the convention for the Riemann tensor is 
$R^\mu{}_{\nu\alpha\beta}\equiv\partial_\alpha\Gamma^\mu_{\nu\beta}
-\partial_\beta\Gamma^\mu_{\nu\alpha}
+\Gamma^\mu_{\lambda\alpha}\Gamma^\lambda_{\nu\beta}
-\Gamma^\mu_{\lambda\beta}\Gamma^\lambda_{\nu\alpha}$.  
Greek letters denote indices in the global coordinate frame, lowercase Latin 
letters from the beginning of the alphabet denote indices in the local Lorentz 
frame, and upper case Latin letters from the beginning of the alphabet denote 
the worldline tetrad frame. All indices run from 0 to 3, while spatial tensor 
indices from 1 to 3, are denoted with lowercase Latin letters from the middle 
of the alphabet. Square brackets on indices denote that they are in the 
worldline tetrad frame. Uppercase Latin letters from the middle of the alphabet 
denote particle labels. The scalar triple product appears here with no brackets, 
i.e.~$\vec{a}\times\vec{b}\cdot\vec{c}\equiv(\vec{a}\times\vec{b})\cdot\vec{c}$.

\section{Setup of EFT for gravitating spinning objects} \label{setup}

\subsection{Setup and goal} \label{eftsetup}

We begin by recalling the general setup of an EFT for the binary inspiral 
problem in terms of a tower of EFTs, building on 
\cite{Goldberger:2004jt,Goldberger:2007hy}.
The binary inspiral problem 
involves two intermediate scales below the radiation wavelength scale, 
$\lambda$, which are the scale of internal structure of each of the compact 
components of the binary, $r_s\sim m$, where $m$ is the mass of the compact 
object, and the orbital separation scale, $r$. It holds that 
$r\sim r_s/v^2 \sim \lambda v$, where $v$ is the typical nonrelativistic 
orbital velocity at the inspiral phase, that is $v\ll 1$. Hence, there is a 
hierarchy of scales in the binary inspiral problem, which makes it ideal for 
an EFT treatment. We note that we consider here gravitating objects, which are in general \textit{spinning}.

Therefore, to obtain an EFT describing the radiation from the binary, one 
should proceed in two stages: 
\begin{enumerate}
\item First, we should have an EFT that removes the scale of the compact 
objects, $r_s$, from the purely gravitational action of the isolated compact
object, which is just the Einstein-Hilbert action
\be\label{sefffull}
S\left[g_{\mu\nu}\right]=-\frac{1}{16\pi G} \int d^4x \sqrt{g} R.
\ee
We integrate out the strong field modes, $g^{s}_{\mu\nu}$, where 
$g_{\mu\nu}\equiv g^{s}_{\mu\nu}+\bar{g}_{\mu\nu}$, by writing down an 
effective action containing the most general set of worldline operators 
consistent with the symmetries of the theory. According to the decoupling 
theorem \cite{Appelquist:1974tg} the effective action can be expressed by 
introducing an infinite tower of worldline operators $O_i(\sigma)$, such 
that
\be\label{seffobject}
S_{\text{eff}}\left[y^\mu,e^{\mu}_{A},\bar{g}_{\mu\nu}\right]=
-\frac{1}{16\pi G} \int d^4x \sqrt{\bar{g}} R\left[\bar{g}_{\mu\nu}\right] + 
\underbrace{\sum_{i}C_i\int d\sigma O_i(\sigma)}_{S_{pp}\equiv
\text{point particle action}},
\ee
where $y^\mu$ and $e^{\mu}_{A}$ are the particle worldline coordinate 
and worldline tetrad degrees of freedom (DOFs), discussed in the following 
section. All UV dependence shows up only in the Wilson coefficients $C_i(r_s)$ 
in the point particle action, $S_{\text{pp}}$, and the worldline operators 
$O_i(\sigma)$ must respect the symmetries of the relevant DOFs at this scale. 
In sections \ref{dof} and \ref{symmetries} below we elaborate 
on the degrees of freedom and the symmetries, considering gravitating 
\textit{spinning} objects. 

We note that a spinning point particle is characterized by two parameters, its 
mass, $m$, and spin length, $S^2$, to be defined in sections 
\ref{eftwithworldlinespin} and \ref{constructnmcs}. Yet, since 
$S\lesssim m^2\sim r_s^2$, then indeed $r_s$ is the only scale in the full 
theory. In addition, dissipative effects from the absorption of gravitational 
waves by the compact objects, as considered in e.g.~\cite{Poisson:2004cw}, 
which modify the mass and spin of the objects, enter only as of the 5PN order. 
Hence, the mass and spin length can be considered as constant for all relevant 
implementations. 
\item The following EFT in the tower should have the orbital scale of the binary 
removed. The metric field is again decomposed into the modes 
\be \label{orbitdecompose}
\bar{g}_{\mu\nu}\equiv\eta_{\mu\nu}+\underbrace{H_{\mu\nu}}_{\text{orbital}}
+\underbrace{\widetilde{h}_{\mu\nu}}_{\text{radiation}}, 
\ee
and we note that 
\be \label{orbitfreq}
\partial_t H_{\mu\nu}\sim \frac{v}{r} H_{\mu\nu}, \qquad 
\partial_\rho\widetilde{h}_{\mu\nu}\sim \frac{v}{r} \widetilde{h}_{\mu\nu}, \ee
whereas 
\be \label{orbitmomentum} \partial_i H_{\mu\nu}\sim \frac{1}{r} H_{\mu\nu}.\ee
This EFT of the binary, which is regarded now as a single composite object, is 
obtained by explicitly integrating out the field modes below the orbital scale, 
$H_{\mu\nu}$. Starting from an effective action of a binary, given by
\be
S_{\text{eff}}\left[y^\mu_{1}, y^\mu_{2}, e_{(1)}{}^{\mu}_{A}, e_{(2)}{}^{\mu}_{A}, 
\bar{g}_{\mu\nu}\right]=
-\frac{1}{16\pi G} \int d^4x \sqrt{\bar{g}} R\left[\bar{g}_{\mu\nu}\right] + 
 S_{(1)\text{pp}}+S_{(2)\text{pp}},
\ee
the effective action of the composite object is defined by the functional integral
\be \label{orbitintegrateout}
e^{iS_{\text{eff(composite)}}\left[y^\mu, e^{\mu}_A,\widetilde{h}_{\mu\nu}\right]} 
\equiv \int {\cal{D}}H_{\mu\nu}~e^{iS_{\text{eff}}\left[y^\mu_{1}, y^\mu_{2},
e_{(1)}{}^{\mu}_{A}, e_{(2)}{}^{\mu}_{A}, \bar{g}_{\mu\nu}\right]}, 
\ee
considering the classical limit, i.e.~evaluating the relevant Feynman diagrams 
in the tree level approximation.
Here $y^\mu$ and $e^{\mu}_A$ are the worldline coordinate and tetrad, 
i.e.~$\eta^{AB}e_A{}^\mu(\sigma)e_B{}^\nu(\sigma)=
\eta^{\mu\nu}+\widetilde{h}^{\mu\nu}$, 
of the composite particle, respectively.
\end{enumerate}
To obtain the final EFT in the tower, an EFT of radiation, the field DOFs should 
all be integrated out. Therefore, in general one has to proceed to a third 
stage, where also the radiation modes, $\widetilde{h}_{\mu\nu}$, are integrated out. 
Yet, in the conservative sector, where no radiation modes are present, and 
from which the conservative dynamics is inferred, the EFT construction 
process ends after the aforementioned two stages, that is after having 
integrated out the field modes below the orbital scale. 

Indeed, in this paper we focus on the imperative two stage process for the 
conservative sector, where the end goal of this process should be an 
\emph{effective} action, i.e.~an action without any remaining orbital scale 
field DOFs. Naturally, this also involves eliminating all unphysical DOFs
from the action, in particular those associated with the rotational DOFs, 
see section 3 in \cite{Levi:2014sba}. 
By definition, e.g.~in eqs.~\eqref{orbitdecompose} and 
\eqref{orbitintegrateout}, an effective action should not contain any 
remaining field DOFs of modes of the scale, which it removes. These should 
all be integrated out. 

It should also be noted that this construction of the EFT, starting 
from the scale of the internal structure of the compact objects, $r_s$, 
should be supplemented below this scale for compact stars, rather than just 
black holes. This becomes relevant, when nonminimal couplings should be 
taken into account, and we comment on that in section \ref{nmcs}. 

\subsection{Degrees of freedom} \label{dof}

We should specify and keep track of our degrees of freedom in the process 
of constructing the EFTs. We should consider here three kinds of DOFs: 
\begin{enumerate}
\item \emph{The gravitational field}. For the effective action in 
eq.~\eqref{seffobject} we 
have the field DOFs in the purely gravitational action, and in the 
non-spinning point particle actions, simply represented by the metric 
$g_{\mu\nu}(x)$ (the overbar notation of the metric is dropped here and 
henceforth).
For the point particle actions beyond the mass monopole, 
which also involve the spins, the tetrad field, 
$\eta^{ab}\tilde{e}_a{}^\mu(x)\tilde{e}_b{}^\nu(x)=g^{\mu\nu}(x)$,
which couples to the multipoles of the objects, also represents the field 
DOFs. 
After gauge fixing the purely gravitational action, and the tetrad field, 
both the metric and the tetrad fields are left with 6 DOFs. 
\item \emph{The particle worldline coordinate}. $y^\mu(\sigma)$ is a 
function of an arbitrary affine parameter $\sigma$. The time coordinate is 
used to fix the gauge of the affine parameter, and we have the 3 DOFs, 
giving the position of the particle. The particle worldline position does not in 
general coincide with the `center' of the object, that is the reference point 
within the actual extended object. The `center' is uniquely defined in Newtonian 
physics, but not in relativity theory. 
\item \emph{The particle worldline rotating DOFs}. Initially, we consider the 
worldline tetrad, an orthonormal frame 
$\eta^{AB}e_A{}^\mu(\sigma)e_B{}^\nu(\sigma)=g^{\mu\nu}$, localized on the 
particle worldline, connecting the body-fixed and general coordinate frames. 
From this tetrad we define the worldline angular velocity 
$\Omega^{\mu\nu}(\sigma)$, and then we add its conjugate, the worldline spin, 
$S_{\mu\nu}(\sigma)$, as a further DOF. We then have $6+6$ DOFs. 
The field DOFs, represented by the tetrad field, which satisfies
$\tilde{e}_a{}^\mu\left(y(\sigma)\right)=\Lambda_a{}^A(\sigma)e_A{}^\mu(\sigma)$, 
are then disentangled from the worldline tetrad DOFs, such that we are left 
with the worldline Lorentz matrices DOFs,
$\eta^{AB}\Lambda_A{}^a(\sigma)\Lambda_B{}^b(\sigma)=\eta^{ab}$, and the conjugate 
worldline spin, $S_{ab}(\sigma)$, projected to the local frame. After gauge 
fixing the rotational DOFs, we are left only with the $3+3$ physical DOFs.
\end{enumerate}

\subsection{Symmetries} \label{symmetries}

The aforementioned degrees of freedom should be coupled in all possible ways 
allowed by the symmetries of the problem in order to construct the effective 
action. The following symmetries should then be considered:
\begin{enumerate}
\item \emph{General coordinate invariance}, and in particular \emph{parity 
invariance} is also included, which holds for macroscopic objects in 
General Relativity, and is relevant for nonminimal couplings in the 
point particle action, see section \ref{nmcs}.  
\item \emph{Worldline reparametrization invariance}. This is used to 
construct the minimal coupling as well as the non minimal coupling parts of 
the point particle action, see sections \ref{eftwithworldlinespin} and 
\ref{nmcs}, respectively.
\item \emph{Internal Lorentz invariance of the local frame field}. We use 
the 3+3 DOFs of local Lorentz transformations to fix the gauge of the tetrad 
field, which in general has 16 DOFs, such that it is represented by the 10 
DOFs of the metric (before gauge fixing), see section \ref{tetradfield}.
\item \emph{SO(3) invariance of the body-fixed spatial triad}, 
$e_{[i]}^{\mu}$, consisting of the 3 spacelike vectors. This follows from 
the 3 rotational DOFs to orient the massive particle in space in the body-fixed frame. 
In consequence, the worldline spin DOFs are SO(3) tensors in the body-fixed 
frame, which is also relevant for nonminimal couplings in the point particle 
action, see section \ref{nmcs}. This is also discussed in section \ref{spinunGF} 
in relation with the \emph{spin gauge invariance}.
\item \emph{Spin gauge invariance}, that is an invariance under the choice 
of a completion of the body-fixed spatial triad through a timelike vector.
This is a gauge of the rotational variables, i.e.~of the worldline tetrad 
and of the worldline spin. 
It is considered in section \ref{spinunGF}, and further discussed in 
section \ref{spinGF}.
\item We assume that the isolated object has no intrinsic permanent 
multipole moments beyond the mass monopole and the spin dipole. 
This is used in sections \ref{eftwithworldlinespin} and \ref{nmcs}.
Permanent multipole moments may be included through constant SO(3) tensors.  
Yet, recall that mass and spin are conserved for isolated objects, but higher 
multipoles are not. 
\end{enumerate}
We stress that time-reversal symmetry is not assumed here, but instead
terms which violate it are shown not to contribute at the considered order, see also section \ref{nmcs}.

\section{Formulation of EFT for spin} \label{formulationeftfors}

\subsection{EFT with the worldline spin as a further DOF} 
\label{eftwithworldlinespin}

First, we briefly review the essential basic definitions as in, e.g.~section 
III of \cite{Levi:2010zu}. We start by considering the worldline tetrad, an 
orthonormal frame $e^\mu_A(\sigma)$, localized on the particle worldline, 
connecting the body-fixed and general coordinate frames, such that 
$\eta^{AB}e^\mu_Ae^\nu_B=g^{\mu\nu}$ with $\eta^{AB}\equiv$diag[1,-1,-1,-1] 
the flat spacetime Minkowski metric. We recall that the reciprocal tetrad 
is defined by $e^{\mu A}\equiv\eta^{AB}e^\mu_B$. The projections of 
any tensor onto the tetrad frame, and the converse projection onto 
the coordinate frame are then defined as, e.g.~for a vector, 
$V_A\equiv e^\mu_A V_\mu$, and $V_\mu\equiv e^A_\mu V_A$, respectively. 

We proceed to define the antisymmetric angular velocity tensor by 
\be \label{Omegadef}
\Omega^{\mu\nu}\equiv e^\mu_A\frac{De^{A\nu}}{D\sigma},
\ee 
where $D/D\sigma$ is the covariant derivative with respect to the worldline 
parameter $\sigma$, and this is a generalization of the flat spacetime 
definition given by 
$\Omega^{ab}\equiv \Lambda^a_A\frac{d\Lambda^{Ab}}{d\sigma}$ 
\cite{Hanson:1974qy,Bailey:1975fe}. Considering the degrees of freedom and 
symmetries of the problem, noted in the previous section, the point particle 
Lagrangian should be a function of the coordinate velocity, 
$u^\mu\equiv dy^\mu/d\sigma$, the angular velocity from 
eq.~\eqref{Omegadef}, and the metric, that is 
$L_{\text{pp}}\left[u^\mu,\Omega^{\mu\nu}, g_{\mu\nu}\right]$, where the 
dependence in the metric is extended beyond minimal coupling to include the 
Riemann tensor and further covariant derivatives.   
The spin is then defined as the conjugate to the angular velocity, i.e.
\be \label{spindef}
S_{\mu\nu}\equiv-2\frac{\partial L}{\partial\Omega^{\mu\nu}}.
\ee 
The minus sign in this definition is chosen to give the correct form in the 
nonrelativistic limit. It is then beneficial to construct the Lagrangian with 
the spin as a further worldline DOF since it makes sense to utilize the 
spin dipole moment, sourcing the gravitons, similar to the mass monopole, as 
a classical source on the worldline. Another advantage is then that the spin 
becomes an independent variational variable, and the equations of motion 
(EOM) of the spin are then directly and conveniently obtained via an 
appropriate variation of the effective action \cite{Levi:2014sba}. 

Therefore, the point particle action from eq.~\eqref{seffobject} can be 
written as
\begin{align}\label{Spp}
S_{\text{pp}} = \int d \sigma \left[ -m \sqrt{u^2}
								- \frac{1}{2} S_{\mu\nu}\Omega^{\mu\nu}
                + L_{\text{SI}}\left[u^{\mu}, S_{\mu\nu}, g_{\mu\nu}
                \left(y^\mu\right)\right]\right],
\end{align} 
where the first two terms are just the point-mass and rotational minimal 
couplings retained from flat spacetime \cite{Hanson:1974qy,Bailey:1975fe}, 
which are inferred from reparametrization invariance. $L_{\text{SI}}$ stands for 
the nonminimal coupling part of the action, which according to the 
symmetries spelled out in section \ref{symmetries} contains only 
spin-induced multipoles, and as will be further illustrated in section 
\ref{nmcs}, only depends on the worldline DOFs $u^{\mu}$ and 
$S_{\mu\nu}$. The conjugate to the 4-velocity $u^{\mu}$ is the \emph{linear} 
momentum, given by
\begin{equation}\label{defp}
p_{\mu} \equiv -\frac{\partial L}{\partial u^{\mu}}.
\end{equation} 
Clearly, it is Lagrangian dependent and is modified as higher multipoles, 
i.e.~nonminimal couplings, are introduced,
and we have 
\be
p^\mu=m \frac{u^{\mu}}{\sqrt{u^2}}+\mathcal{O}(S^2).
\ee
For an isolated compact object the linear momentum $p^\mu$ can be obtained from
surface integrals at spatial infinity. In this case finite size effects are not taken 
into account, and the mass is matched as $m^2 = p_\mu p^\mu$.

We note that we can also express the rotational minimal coupling term, using 
the spin projected to the body-fixed frame, where the spin is a permanent 
multipole moment. Indeed, the components of the spin in this frame are
constant, which can be seen most directly using the EOM following from the action in 
eq.~\eqref{Spp}. Using the Ricci rotation coefficients, defined by
\be \label{Riccirotation}
\omega_{\mu}{}^{ab} \equiv e^b{}_{\nu} D_{\mu}e^{a\nu},
\ee 
it holds that
\begin{align} \label{mcs}
\frac{1}{2} S_{\mu\nu}\Omega^{\mu\nu}&= 
\frac{1}{2} S_{AB} \omega_{\mu}^{AB} u^{\mu}.
\end{align}
We shall see that only the spatial SO(3) components in the body-fixed frame 
are non-vanishing here.
Considering the scalar mass monopole from eq.~\eqref{Spp}, and this form, 
where the spin dipole is also represented as a constant antisymmetric SO(3) 
tensor, we shall be able to construct the nonminimal coupling part 
of the action in a rather straightforward manner, as will be detailed in 
section \ref{nmcs}.

As we are working in an action approach, there is no impediment to the 
implementation of gauge constraints on the rotational DOFs. Moreover, as we 
shall see these gauge constraints should be implemented at the level of the 
point particle action in order to ultimately arrive at an effective action 
without any remaining orbital scale field degrees of freedom. 
We shall also see in the following section \ref{spinunGF}, that in order to 
arrive at a generic point particle action, where the gauge of the rotational 
variables is not fixed, we should initially implement the covariant gauge. 
Yet it is crucial to point out that in the point particle Lagrangian in 
eq.~\eqref{Spp}, we have both DOFs of the angular velocity and of the spin, and 
therefore it is necessary to implement gauge fixing both on the worldline 
tetrad DOFs and on the spin DOFs, rather than only on the latter ones. We 
shall explicitly see in section \ref{spinGF}, that we cannot obtain an 
effective action formulated with the worldline spin, if the gauge of the 
spin is fixed without gauge fixing its conjugate DOFs. These are principal 
statements in this paper. 

\subsection{Unfixing the gauge of the rotational variables} \label{spinunGF} 

As we noted in section \ref{symmetries} there is a spin gauge 
freedom in the choice of a timelike vector for the worldline tetrad. This 
is a choice of a `center' point within the \textit{spinning} object, which 
must have a finite size due to its spin. This gauge is fixed using some spin 
supplementary conditions (SSC), corresponding to a gauge choice of the 
timelike basis vector for the worldline tetrad. The covariant SSC by Tulczyjew 
\cite{Tulczyjew:1959c}, given by 
\be
S_{\mu\nu} p^{\nu} = 0, 
\ee
where $p^{\nu}$ is the linear momentum from eq.~\eqref{defp}, is the only SSC, 
for which existence and uniqueness of a corresponding `center' were proven rigorously 
in general relativity \cite{Schattner:1979vp,Schattner:1979vn}. 
Together with the corresponding gauge condition on the worldline tetrad 
timelike basis vector, which reads 
\be 
e_{[0]}^\mu = p^\mu / \sqrt{p^2},
\ee
we see that this gauge is equivalent to the requirement that the spin in the 
body-fixed frame, as in eq.~\eqref{mcs}, is a spatial SO(3) tensor.

General coordinate invariance is an important symmetry of the  
effective action, and it makes sense to specify the point particle action 
for this covariant SSC. However, we also expect to have a spin gauge 
symmetry in the effective action, such that indeed there is gauge 
freedom in the choice of the timelike vector for the worldline tetrad. This 
gauge symmetry was not addressed in previous work, which approached to extend the EFT formulation for spinning 
objects. Therefore, we start out with the covariant gauge, having a covariant 
theory at hand, and then transform to a generic gauge, thus introducing 
the spin gauge freedom into the action. Hence, we illustrate a procedure for 
directly constructing an action, which is manifestly invariant under both 
coordinate and rotational variables gauge transformations.

We are approaching the change of a spin gauge from a new perspective. We are 
effectively applying a boost to the body-fixed tetrad, and then we see how the 
rotational minimal coupling term $\frac{1}{2} S_{\mu\nu}\Omega^{\mu\nu}$ in 
the action in eq.~\eqref{Spp} is affected.  
This approach is in fact suggested by the EFT philosophy since an important 
ingredient in the EFT setup is the SO(3) invariance of the worldline 
spatial triad, rather than SO(1,3) Lorentz invariance of the worldline tetrad. 
Since the velocity of the particle is already set as the time derivative of 
its position coordinate, the additional boost degrees of freedom of the 
internal SO(1,3) indices of the worldline tetrad are actually redundant gauge 
ones. Only the orientation of the particle is physical, and can be described 
by an internal SO(3) group. Then the action should be formulated in terms of 
$e_{[i]}{}^{\mu}$, which possess SO(3) indices $i$, without the timelike 
basis vector $e_{[0]}{}^{\mu}$. Hence we have the idea that a fixation of 
$e_{[0]}{}^{\mu}$ should be connected to the gauge choice of the spin 
variable or SSC. We are going to show that this is indeed the case. We 
therefore connect different choices for $e_{[0]}{}^{\mu}$ by effectively 
boosting $e_{A}{}^{\mu}$ in the following. 

Consider a boost in the 4-dimensional covariant form. It is given by
\begin{equation}\label{stdboost}
L^a{}_b(q, w) \equiv \delta^a{}_b + 2 q^a w_b 
- \frac{(q^a + w^a)(q_b + w_b)}{1 + qw} ,
\end{equation}
where $qw \equiv q_a w^a$, and $q^a q_a = w^a w_a=1$, i.e.~$q_a$, $w_a$ are 
timelike unit 4-vectors. 
We will now make use of the definition in eq.~\eqref{stdboost} with general 
coordinate indices instead of Lorentz indices. Then $L^{\mu}{}_{\nu}(q, w)$ is 
strictly speaking not a Lorentz transformation, but it is taken as the tensor 
projected by the appropriate tetrad onto the coordinate frame, see
e.g.~\cite{Fahnline:1983}.
We can then transform $e^{A\mu}$ from a gauge condition
\begin{equation} \label{oldgauge}
e_{A\mu} q^{\mu}=\eta_{[0]A}  \Leftrightarrow e_{[0]\mu} = q_{\mu},
\end{equation}
to the condition
\begin{equation} \label{newgauge}
\hat{e}_{A\mu} w^{\mu}=\eta_{[0]A} \Leftrightarrow \hat{e}_{[0]\mu} = w_{\mu},
\end{equation}
with the help of the transformation
\begin{equation}
\hat{e}^{A\mu} = L^{\mu}{}_{\nu}(w, q) e^{A\nu} .
\end{equation}
The algebraic properties of $L^{\mu}{}_{\nu}(q, w)$ then guarantee that
\begin{equation}
\eta_{AB} \hat{e}^{A\mu} \hat{e}^{B\nu} = g^{\mu\nu},
\end{equation}
if the analogous relation holds for $e^{A\mu}$.

Further, notice that from eqs.~\eqref{oldgauge} and \eqref{newgauge} we have
\begin{align}
\frac{D q^{\mu}}{D \sigma} &= - \Omega^{\mu\nu} q_{\nu} , \\
\frac{D w^{\mu}}{D \sigma} &= - \hat{\Omega}^{\mu\nu} w_{\nu} ,
\end{align}
where similarly to eq.~\eqref{Omegadef}, we have 
$\hat{\Omega}^{\mu\nu}\equiv\hat{e}_A{}^{\mu} \frac{D\hat{e}^{A\nu}}{D\sigma}$.
For the angular velocity we get
\begin{equation}
\Omega^{\mu\nu} = L^{\mu}{}_{\rho}(q, w) L^{\nu}{}_{\sigma}(q, w) 
                  \hat{\Omega}^{\rho\sigma}
	                + L^{\mu}{}_{\rho}(q, w) \frac{D L^{\nu\rho}(q, w)}{D \sigma},
\end{equation}
and since it holds that
\begin{equation}
L^{\mu}{}_{\rho}(q, w) \frac{D L^{\nu\rho}(q, w)}{D \sigma} =
	2 \frac{D w^{\mu}}{D \sigma} q^{\nu}
	- \frac{2 w^{\mu} q^{\nu} q_{\rho}}{1+qw} \frac{D w^{\rho}}{D \sigma}
	- \frac{q^{\nu} + w^{\nu}}{1+qw} \frac{D (q^{\mu} + w^{\mu})}{D \sigma}
	- (\mu \leftrightarrow \nu),
\end{equation}
then the transformation between the covariant angular velocities finally 
reads
\begin{equation}
\Omega^{\mu\nu} = \hat{\Omega}^{\mu\nu}
+ \left[ \frac{q^{\mu} + w^{\mu}}{1+qw} \left( \frac{D q^{\nu}}{D \sigma} + 
\hat{\Omega}^{\nu\rho} q_{\rho} \right) - (\mu \leftrightarrow \nu) \right].
\end{equation}
From this we can easily obtain the effect on the minimal coupling term as
\begin{equation}\label{SmcTrans}
\frac{1}{2} S_{\mu\nu} \Omega^{\mu\nu} =
	\frac{1}{2} \left( S_{\mu\nu} - \frac{S_{\mu\rho} w^{\rho}}{1+qw} q_{\nu}
		+ \frac{S_{\nu\rho} w^{\rho}}{1+qw} q_{\mu} \right) \hat{\Omega}^{\mu\nu}
	- \frac{S_{\mu\rho} w^{\rho}}{1+qw} \frac{D q^{\mu}}{D \sigma},
\end{equation}
where the SSC $S_{\mu\nu} q^{\nu} = 0$ was used. Notice that both the gauge 
condition on the tetrad, $e_{[0]\mu} = q_{\mu}$, and the SSC should be used 
at the level of the action in order to transform the minimal coupling term.

The last equation suggests to define a new spin tensor as the prefactor of 
the transformed angular velocity $\hat{\Omega}^{\mu\nu}$, that is
\begin{equation}
\hat{S}_{\mu\nu} \equiv S_{\mu\nu} - \frac{S_{\mu\rho} w^{\rho}}{1+qw} q_{\nu}
		+ \frac{S_{\nu\rho} w^{\rho}}{1+qw} q_{\mu} .
\end{equation}
If indeed we fix $q^{\mu} = p^{\mu} / \sqrt{p^2}$, i.e.~we start with 
Tulczyjew's covariant SSC, then the new spin variable reads 
\begin{equation}\label{TulczyjewSTrans}
\hat{S}_{\mu\nu}=S_{\mu\nu} -\frac{S_{\mu\rho} w^{\rho}}{\sqrt{p^2}+ p w}p_\nu
	+ \frac{S_{\nu\rho} w^{\rho}}{\sqrt{p^2} + p w} p_{\mu},
\end{equation}
and satisfies the generic spin supplementary condition
\begin{equation}\label{genSSC}
\hat{S}^{\mu\nu} \left( p_{\nu} + \sqrt{p^2} \hat{e}_{[0]\nu} \right) = 0,
\end{equation}
together with the gauge constraint for the tetrad
\be \label{gengauge}
\hat{e}_{[0]\mu} = w_{\mu},
\ee
which provide together the necessary 3+3 gauge constraints to eliminate the redundant 
unphysical DOFs in the angular velocity and spin tensors, see section \ref{spinGF} 
for further analysis of specific sensible gauges.

Notice also that the spin transforms as expected under a shift of the `center'
position, $\delta z^{\mu} \equiv \hat{z}^{\mu} - y^{\mu}$, namely as
\begin{equation}\label{genSTrans}
\hat{S}^{\mu\nu} = S^{\mu\nu} - \delta z^{\mu} p^{\nu} + \delta z^{\nu} p^{\mu}.
\end{equation}
By comparing eqs.~\eqref{genSTrans} and \eqref{TulczyjewSTrans} we can read 
off $\delta z^{\mu}$ as
\begin{equation} \label{genworldline}
\delta z^{\mu} = \frac{S^{\mu\rho} w_{\rho}}{\sqrt{p^2} + p w} .
\end{equation}
Notice that due to the initial covariant SSC it holds that 
$\delta z^{\mu} p_{\mu} = 0$, that is the shift of the `center' of the object 
is orthogonal to the linear momentum, and thus indeed spacelike.
Hence, the `center' of the spinning particle is shifted from the worldline 
position of the particle for a non-covariant gauge of the particle rotating 
DOFs.

Therefore we see that by choosing a gauge for the tetrad in eq.~\eqref{gengauge}, 
we also specify the choice of SSC in eq.~\eqref{genSSC}, the spin variable in  
eq.~\eqref{TulczyjewSTrans}, and the position of the `center' in
eq.~\eqref{genworldline}. 
At this point we have arrived at
\begin{align}\label{mcTrans}
\frac{1}{2} S_{\mu\nu} \Omega^{\mu\nu} &= 
  \frac{1}{2} \hat{S}_{\mu\nu} \hat{\Omega}^{\mu\nu}
	- \delta z^{\mu} \frac{D p_{\mu}}{D \sigma}.
\end{align}
Thus, we have introduced spin gauge freedom into the rotational minimal 
coupling term of the point particle action in eq.~\eqref{Spp}. An 
alternative construction of the spin gauge symmetry in flat spacetime, 
based on the canonical formalism, is given in \cite{Steinhoff:2015ksa}.
 
In order to also introduce this gauge freedom beyond minimal coupling 
according to eq.~\eqref{Spp}, see section \ref{nmcs}, we need to express our 
initial spin variable in terms of the generic one in 
eq.~\eqref{TulczyjewSTrans}. We recall that we consider that also the nonminimal 
coupling terms have been initially constructed for a covariant gauge with 
a spin variable, satisfying Tulczyjew's covariant SSC. Then we note that from the 
contraction of eq.~\eqref{TulczyjewSTrans} with 
$p^{\nu}$, we get that the shift of position of the `center' can be written as    
\begin{equation} \label{covariantshift}
\delta z^\mu= \frac{S^{\mu\rho} w_{\rho}}{\sqrt{p^2} + p w} 
= - \frac{\hat{S}^{\mu\rho} p_{\rho}}{p^2},
\end{equation}
which leads to
\begin{equation}\label{Strans}
S_{\mu\nu} = \hat{S}_{\mu\nu} - \frac{\hat{S}_{\mu\rho} p^{\rho} p_{\nu}}{p^2}
	+ \frac{\hat{S}_{\nu\rho} p^{\rho} p_{\mu}}{p^2}.
\end{equation}
The transformation to the new generic spin can therefore be written without the 
$w_{\mu}$ gauge DOFs. This is so since eq.~\eqref{Strans} is a projection of the 
generic spin variable $\hat{S}_{\mu\nu}$ onto the spatial hypersurface of the rest 
frame. This projection removes the gauge DOFs, since all gauges agree in the rest 
frame, hence the projected spin variable is spin gauge invariant. It follows that 
indeed the action should be constructed using $S_{\mu\nu}$. Recall that the spin 
variable, which satisfies the covariant SSC, when projected to the body-fixed frame, 
is a spatial constant SO(3) tensor. 

Finally, the transformation to the generic worldline triad reads
\begin{equation}\label{etrans}
e_{[i]}{}^{\mu} = \hat{e}_{[i]}{}^{\mu}
        - \hat{e}_{[i]}{}^{\rho} p_{\rho} \frac{p^{\mu} + \sqrt{p^2} 
        \hat{e}_{[0]}{}^{\mu}}{p^2 + \sqrt{p^2} p_{\nu} \hat{e}_{[0]}{}^{\nu}},
\end{equation}
where we have for the temporal component that 
$e_{[0]}{}^{\mu} = p^{\mu} / \sqrt{p^2}$, and $e_{[i]}{}^{\mu} p_{\mu} = 0$.
Notice that only in this relation the new worldline tetrad time vector
$\hat{e}_{[0]}{}^{\nu}=w_{\mu}$ appears explicitly. Hence, the point particle 
action beyond minimal coupling from eq.~\eqref{Spp} is constructed from 
$S_{\mu\nu}$ and $e_{[i]}{}^{\mu}$, but these are now understood in terms of 
eqs.~\eqref{Strans} and \eqref{etrans}. Actually, eq.~\eqref{etrans} is 
not required for spin interactions beyond minimal coupling, but it is 
needed for dynamical tidal interactions. 

Finally, the worldline variables and fields in the point particle action are still taken at 
$y^{\mu}$. That is, for a non-covariant gauge the worldline position would be shifted 
from the location $\hat{z}^\mu$ of the `center', which one commonly attributes to 
$\hat{S}_{\mu\nu}$. 
Instead, if one chooses to eliminate the covariant derivative in 
eq.~\eqref{mcTrans} at the level of the point particle action by making a shift of the 
worldline to the location of the `center', then the worldline variables and fields are 
implicitly parallel transported to $\hat{z}^{\mu}$. Yet, this is not judicious in order 
to integrate out all the orbital field modes from the effective action, which can be 
obtained only if the rotational gauge fixing is implemented at the level of the point 
particle action. Further, at this stage it is preferable to keep the point particle 
action in its general form. For further discussion on this point, see also next 
section \ref{etmc}, and section \ref{spinGF}. 

\paragraph{Using the coordinate velocity in the rotational gauge fixing.} 

So far, we have made all derivations regarding the rotational gauge fixing 
in terms of the linear momentum $p_{\mu}$ for generality. Yet, from 
eqs.~\eqref{Spp} and \eqref{defp} we can tell, that the difference 
$p_{\mu}-mu_{\mu}/u$ is quadratic in the spin, and linear in the field. Then 
from Tulczyjew SSC, $S_{\mu\nu}p^\nu=0$, we can deduce, that this difference 
may be relevant only as of cubic order in the spin, and in that case, only as 
of NLO, due to the resulting nonlinear coupling of the field to spin. 
Therefore, for all current cases of interest, up to the 4PN order for rapidly 
rotating compact binaries, we can replace the linear momentum in Tulczyjew 
SSC with its LO coordinate velocity approximation. In particular, note that this 
applies to the useful eqs.~\eqref{Strans}, \eqref{genSSC}, \eqref{mcTrans}, 
\eqref{etrans}, which change the spin variable, fix the redundant temporal 
spin components, $S^{0i}$, transform the rotational minimal coupling term, 
and the worldline tetrad, respectively. 

\subsection{Extra term from minimal coupling} \label{etmc}

Let us now focus on the second term on the right hand side of eq.~\eqref{mcTrans}, 
which involves a covariant derivative of the linear momentum, that is
\be \label{mcextra}
\frac{\hat{S}^{\mu\nu} p_{\nu}}{p^2} \frac{D p_{\mu}}{D \sigma}=
- \delta z^{\mu} \frac{D p_{\mu}}{D \sigma}.
\ee
This term essentially adds the Thomas precession of the spin in curved spacetime. 
In flat spacetime the spin does not precess in the absence of a torque, and this term 
essentially does not contribute. In curved spacetime there is a torque: the force of 
curvilinear motion on the worldline due to the gravitational field operating on the 
`arm', which is the shift from the worldline to the `center'.
Hence, in curved spacetime this term is relevant, and it should be taken into 
account to all orders in spin. If the covariant derivative of the momentum is 
eliminated, using the EOM at the level of the point particle action, which corresponds 
to a redefinition of the position \cite{Damour:1990jh,Levi:2014sba}, it 
contributes once spin dipole effects are taken into account in the EOM, and thus from 
eq.~\eqref{mcextra} we see that it first appears in the NLO spin-squared 
sector. That is, it manifestly contributes as a finite size effect once the spin of 
the `particle' is taken into account, as a spinning `particle' cannot actually be 
considered a `particle' anymore, but instead it must have an extended finite size, 
hence its `arm' $\sim S/m$, where S is the spin length, defined in eq.~\eqref{sldef}. 

It is important to point out the relation of our results to the work by Yee 
and Bander in \cite{Yee:1993ya}, which advocated a Routhian approach for the 
obtainment of the EOM of the spin, and considered up to and including the  
quadrupolar level. If we consider our eqs.~\eqref{Strans} for the spin 
variable, and eq.~\eqref{mcextra} for the extra term from minimal coupling, we 
can see that they are similar to eqs.~(7) and (9) of \cite{Yee:1993ya}, where 
their corresponding terms are considered only in the local Lorentz frames, and 
in the approximation $p_\mu\simeq m u^\mu/u$. 

Yet, it is necessary to make the significant distinction between our 
formulation here, and their related procedure. Yee and Bander present the 
change of the spin variable, and the addition of the extra term as an 
\textit{ad hoc} procedure to ensure the covariant SSC is satisfied, and to 
obtain the EOM of the spin via the Poisson brackets, while staying at the so(1,3) 
level, and not actually implementing the covariant SSC. The work in 
\cite{Porto:2008tb, Porto:2008jj} followed their procedure 
for the treatment of spin, but substituted the worldline acceleration in 
the extra term similar to eq.~\eqref{mcextra}, using the Mathisson-Papapetrou EOM 
\cite{Mathisson:1937zz, Mathisson:2010, Papapetrou:1951pa} for a 
pole-dipole approximation, given by
\begin{align}
\frac{D p_{\mu}}{D \sigma}=-\frac{1}{2} R_{\mu\nu\rho\kappa}u^{\nu} 
S^{\rho\kappa}. 
\end{align}
This substitution is a linear approximation in the shift of the worldline, where 
terms beyond linear in the shift were dropped from the action via the application 
of the covariant SSC. Thus, an extra Riemann dependent term was introduced, 
linear in the covariant SSC, which contributes to the NLO spin-squared sector. 

First, here we see that since this term originates from minimal coupling, it actually 
carries no Wilson coefficient. More importantly, it should be stressed that there is 
no need to require the preservation of the SSC in the action, and restrict it to the 
covariant one. 
The gauge constraints on the worldline tetrad and the SSC should be implemented at 
the level of the point particle action as was put forward in \cite{Levi:2008nh}. 
Then, using the internal symmetry of the worldline tetrad to incorporate the spin 
gauge freedom in the point particle action, we obtain here an extra term 
from minimal coupling. 
Moreover, here we shall obtain an effective action with the physical SO(3) rotational 
variables, and where all orbital field modes are absent. We then conveniently obtain 
the EOM of the spin via a proper variation of the action, which take on a very simple 
form, see section \ref{spineom}.

\section{EFT for spin: Nonminimal couplings} \label{nmcs}

In this section we proceed to construct the point particle action for the 
spinning particle beyond minimal coupling, which accounts for the finite size 
effects of the gravitating object. These nonminimal couplings carry Wilson 
coefficients, which encode the internal structure of the object, and should 
be found by matching the effective theory with the full theory. Here we 
consider nonminimal couplings of the self-induced multipoles from spin, which 
contribute as of the 2PN order for rapidly rotating compact objects. Tidal 
nonminimal couplings, as considered in \cite{Poisson:2004cw}, contribute only as 
of the 5PN order, and thus are not of interest for current implementations.
It should be noted however that the Wilson coefficients of tidal nonminimal 
couplings can be large, and tidal interactions are also expected to be 
important for gravitational wave astronomy.

We first spell out the general considerations for the construction of these 
nonminimal couplings, building on \cite{Goldberger:2004jt, Goldberger:2005cd, 
Goldberger:2007hy, Goldberger:2009qd}. The point particle action in 
eq.~\eqref{Spp} is augmented with higher order operators, where derivatives of 
the field are added along with higher multipoles. Hence, these new terms are 
suppressed by powers of the ratio of the scale of the source to the orbital 
separation, namely they are naturally ordered by their PN relevance. Then, we 
present the LO nonminimal couplings to all orders in the spin. These were already 
implemented in \cite{Levi:2014gsa} by means of the EFT formulation we present here, where 
the cubic and quartic in spin sectors were fully obtained for generic compact 
binaries. 

As we consider the nonminimal couplings, where the internal structure of the 
objects starts to play a role, let us comment further on the physics of 
actual astrophysical extended objects. For generic stars the tower of EFTs 
presented in section \ref{eftsetup} starts below the scale $r_s$, from a 
fluid description of matter down to elementary particle interactions. That 
is, one must add an EFT for matter to eq.~(\ref{sefffull}), e.g.~an ideal 
fluid action. The scale of the object can be integrated out explicitly for 
idealized Newtonian stars \cite{Chakrabarti:2013xza}, which leads to a 
point-mass action augmented by harmonic oscillator DOFs, corresponding to 
oscillation modes of the star, which couple to the tidal forces. This is 
found to approximately hold also for stars in general relativity 
\cite{Chakrabarti:2013lua}. It should be noted that when the orbital 
frequency is in resonance, an infinite number of terms would be needed in 
eq.~(\ref{seffobject}), which indicates that further DOFs should be added 
to the EFT. This is analogous to resonances in EFTs in particle physics 
\cite{Georgi:1994qn}, and was overlooked in 
\cite{Goldberger:2004jt, Goldberger:2005cd}.

\subsection{Construction of spin nonminimal couplings} \label{constructnmcs}

We start by discussing the possible spin-induced multipoles. 
First, from section \ref{symmetries} we recall that the action should be parity 
invariant. Every operator must therefore 
contain an even number of odd parity tensors, i.e.~of the Levi-Civita tensor
$\epsilon_{\alpha\beta\gamma\lambda}=\sqrt{-g}[\alpha\beta\gamma\lambda]$, 
where $g$ is the determinant of $g_{\mu\nu}$, and $[\alpha\beta\gamma\lambda]$ 
is the totally antisymmetric Levi-Civita symbol with $[0123]=+1$. Therefore, 
we should consider possible dual tensors, in particular the dual of the spin 
tensor, given by
\begin{equation}
*\!S_{\alpha\beta}\equiv\frac{1}{2} \epsilon_{\alpha\beta\mu\nu} S^{\mu\nu}.
\end{equation}
Its contraction with the 4-velocity is the spin vector, given by
\be \label{svec}
S^{\mu}\equiv *S^{\mu\nu} \frac{p_{\nu}} {\sqrt{p^2}} 
\simeq *S^{\mu\nu} \frac{u_{\nu}} {\sqrt{u^2}}.
\ee
Due to the orthogonality of the spin tensor $S^{\mu\nu}$ to $u_{\nu}$ from the 
covariant SSC, we shall see that the spin vector $S^{\mu}$ is the only 
combination, with which $u_{\nu}$ can enter the spin-induced multipoles. 
Notice that $S^{\mu} u_{\mu} = 0$, 
i.e.~$S^{\mu}$ is a spacelike vector.
Also note that the inverse of the dual is 
$*\!*\!S^{\alpha\beta} = -S^{\alpha\beta}$.
Using the SSC, i.e.~the orthogonality of $S^{\mu\nu}$ to $u_{\mu}$, at the 
level of the action, we have then
\begin{align}
S_{\mu\alpha} *\!S^{\alpha\nu} =& 
- \frac{1}{4} \delta^{\nu}_{\mu} S_{\alpha\beta} *\!S^{\alpha\beta} = 0, \\
S_{\mu\alpha} S^{\alpha} =& \,0 , \\
S_{\mu} S^{\mu} =& - \frac{1}{2} S_{\mu\nu} S^{\mu\nu}\equiv -S^2, \label{sldef}
\end{align}
see e.g.~\cite{Hanson:1974qy} for similar identities in flat spacetime, and 
we defined the spin length $S^2$ with the minus sign from the spacelike spin 
vector.

Now, from the Cayley-Hamilton theorem we expect that higher powers of the 
spin tensor are dependent. Then, let us examine higher powers of the spin 
tensor in the sense of matrix multiplication. We have 
\begin{align}
S^{\alpha}{}_{\mu} S^{\mu}{}_{\beta} &= -S^{\alpha} S_{\beta}
-S^2\left(\delta^{\alpha}{}_{\beta}-\frac{u^{\alpha} u_{\beta}}{u^2} \right),
\label{ssquare}\\
S^{\alpha}{}_{\mu} S^{\mu}{}_{\nu} S^{\nu}{}_{\beta}&= -S^2 S^{\alpha}{}_{\beta},
\end{align}
where we also note that the square of the spin is symmetric. Indeed, we see 
that from the last relation we can read off the minimal polynomial of the 
spin matrix $S^{\alpha}{}_{\mu}$ as 
\be \label{eigen}
X(X+iS)(X-iS)=0,
\ee
which is of degree 3, that is lower than the degree of the characteristic 
polynomial (4). That means one of the eigenvalues of $S^{\alpha}{}_{\mu}$ is 
degenerate. Indeed, for the eigenvalue $0$ we have both 
$S^{\alpha}{}_{\mu} u^{\mu} = 0$, and $S^{\alpha}{}_{\mu} S^{\mu} = 0$. 
The two remaining eigenvalues of $S^{\alpha}{}_{\mu}$ are the pure imaginary values 
$\pm iS$, as expected for the antisymmetric spin matrix.
Therefore, the determinant of the spin tensor is zero. 
A similar analysis for the dual spin tensor leads to no further independent 
contractions as expected. 

In conclusion, from the above analysis we see that the independent 
combinations of spin that we can use to construct the operators are just the 
even-parity spin tensor $S^{\mu\nu}$, the odd-parity spin vector $S^{\mu}$, 
and the even-parity square of the spin tensor 
$S^{\alpha}{}_{\mu} S^{\mu}{}_{\beta}$, where these spin tensors should not 
be contracted among themselves. 

It should be stressed that we implement the covariant SSC at the level of the 
point particle action, which reduces the number of possible spin-induced 
multipoles. Yet, since ultimately the spin-induced multipoles should be 
expressed in terms of the gauge invariant projected spin in eq.~\eqref{Strans}, 
which is algebraically orthogonal to $u^{\mu}$, then these possibilities would 
be omitted for all gauge spin variables.

Further, from section \ref{symmetries} we recall the SO(3) invariance of the 
body-fixed frame triad. Since we consider our spinning extended body to be a 
point particle, then there exists a locally inertial frame,
which is comoving and in which the spin does not precess \cite{Weinberg:1972}.
Yet, finite size effects lead to spin precession in this frame.
From this comoving, approximately inertial frame, one can move to a corotating
one, that is to the body-fixed frame. In the latter, the spin components are
constant even in the presence of finite size effects. Moreover, in this locally 
flat frame we have the SO(3) invariance as Wigner little group, therefore 
tensors form SO(3) irreducible representations \cite{Weinberg:1995mt}. Indeed, 
in agreement with our analysis of the spin tensor in matrix representation, we 
find from eq.~\eqref{eigen}, that the spin tensor transforms as a massive vector 
particle, i.e.~like a vector in space. We recall from eqs.~\eqref{Spp} and \eqref{mcs}, 
that the minimal coupling part of the action is given with the scalar mass monopole, 
and the spin dipole, represented by constant SO(3) tensors in the body-fixed frame, 
where the antisymmetric spin is also spatial in the covariant SSC. Hence, the 
spin-induced higher multipoles should naturally be considered in the body-fixed frame.

Just like in the minimal coupling part of the action, we start from the covariant gauge, where we have for the body-fixed tetrad that 
$e_{[0]}{}^{\mu} = u^{\mu}/\sqrt{u^2}$, and $e_{[i]}{}^{\mu} u_{\mu} = 0$. 
Then it is easy to see that in the body-fixed frame it holds for the spin vector from eq.~\eqref{svec}, that $S_{[0]} = 0$, and for the spatial components 
\begin{equation}
S^{[i]} = \frac{1}{2} \epsilon^{[ijk]} S^{[jk]},
\end{equation}
where this actually holds in every locally flat frame, and the indices are just Euclidean here. Therefore, the spin vector is also constant in the body-fixed frame. Hence, the spin-induced higher multipoles are symmetric traceless constant spatial tensors in the body-fixed frame. It should be noted that the constant scalar spin length $S^2$, which is the trace of the square of the spin tensor, is absorbed in the mass, and the Wilson coefficients, and is therefore omitted from these traceless tensors. 

These even and odd spin-induced multipoles couple to the even and odd 
parity electric and magnetic curvature tensors, respectively, and their 
covariant derivatives. The electric and magnetic curvature tensors are usually defined 
with the Weyl tensor, yet the field here is a vacuum solution at 
LO, hence Ricci tensor terms can be made to vanish using field 
redefinitions, and so the use of Weyl and Riemann tensors is equivalent. Then, 
we define the electric component of the Riemann tensor as 
\be \label{Edef}
E_{\mu\nu}\equiv R_{\mu\alpha\nu\beta} u^{\alpha}u^{\beta},
\ee
and the magnetic component of the Riemann tensor as
\be \label{Bdef}
B_{\mu\nu}\equiv\frac{1}{2}\epsilon_{\alpha\beta\gamma\mu} R^{\alpha\beta}_
{\quad\delta\nu} u^{\gamma}u^{\delta},
\ee
where here the dual of the Riemann tensor $*\!R_{\gamma\mu\delta\nu}\equiv
\frac{1}{2}\epsilon_{\alpha\beta\gamma\mu} R^{\alpha\beta}_{\quad\delta\nu}$
is used. In the current work, we consider only couplings linear in Riemann, that is as we noted we are 
not concerned with the tidal response to external gravitational fields, 
which does not contribute at the PN orders of interest.

From the definitions in eqs.~\eqref{Edef}, \eqref{Bdef}, one obtains that both the 
electric and magnetic components of the Riemann tensor are symmetric, 
traceless, and orthogonal to the 4-velocity, using the symmetries of the Riemann tensor, the first Bianchi identity, and their being a vacuum field solution.
These SO(3) tensors are then also considered in the body-fixed frame, where only their projection on the spatial triad is 
non-vanishing due to the covariant gauge of the tetrad. It follows then that 
they are symmetric and traceless also with respect to their internal spatial 
indices. 

Next, we also consider the covariant derivatives of the electric and magnetic 
tensors. These are also projected to the body-fixed frame, 
i.e.~$D_{[i]}=e^{\mu}_{[i]}D_{\mu}$, where we clarify that the covariant 
derivative shall not act on the 4-velocity, contained in $E_{\mu\nu}$ and $B_{\mu\nu}$, 
since it is a function of the worldline parameter $\sigma$ only. 
As for the time derivative
$D_{[0]}=u^{\mu}D_{\mu}\equiv D/D\sigma$, it is just the covariant 
derivative along the worldline.  
Now, at linear order in the curvature time derivatives of the curvature can be 
integrated by parts to time derivatives of the particle variables. Terms 
including such higher order time derivatives of the worldline variables can 
be removed via variable redefinitions with a shift of, e.g.~the worldline 
coordinate, using lower order EOM, and get absorbed into other Riemann 
dependent finite size operators without higher order time derivatives of 
the worldline variables, namely into their Wilson coefficients.
Therefore, we can consider here only the spatial derivatives, projected 
orthogonally to $u^{\mu}$.

The indices of the covariant derivatives are also symmetrized among themselves, 
and with respect to the indices of the electric and magnetic tensors. The first 
symmetrization follows since the commutation of covariant derivatives leads to 
further curvature terms, and as only terms linear in the curvature are considered 
here, such contributions can be taken to vanish. The second symmetrization with 
indices from the covariant derivatives, and from the electric and magnetic components, 
follows from the differential Bianchi identity of the Weyl tensor in vacuum, which 
leads to equations analogous to Maxwell's:
\begin{align}\label{Maxwell}
	\epsilon_{[ikl]} D_{[k]} E_{[lj]} =& \dot{B}_{[ij]}, \\
	\epsilon_{[ikl]} D_{[k]} B_{[lj]}=&  - \dot{E}_{[ij]}.
\end{align}
Notice that the left hand side contains the commutator of derivative and
curvature components indices.
Since as was explained time derivatives of the curvature can be ignored at linear 
order in the curvature, symmetrization follows.
Actually, the indices of the covariant derivatives and of the electric and magnetic 
tensors would anyway be symmetrized here upon contraction with the symmetric 
spin-induced multipoles. It is also clear from the above discussion, that also in the 
generic case, where tidal effects are taken into account, the independent curvature 
tensors are taken as the spatial derivatives of $E_{[ij]}$ and $B_{[ij]}$ with all 
indices symmetrized, and their time derivatives, as in \cite{Bini:2012gu}.

Finally, from further contracting from eq.~\eqref{Maxwell} we also get
\be
D_{[i]} E_{[ij]}=D_{[i]} B_{[ij]}=0,
\ee 
analogous to Maxwell's equations, and so we also have similarly
\be 
\Box E_{[ij]}=\Box B_{[ij]}=0.
\ee 
Hence, these tensors with covariant derivatives are also traceless. 
Therefore, the indices of the curvature components, and of their covariant 
derivatives, should also not be contracted among themselves. 

At this stage it becomes clear how to use these building blocks, which we 
have detailed, to construct the nonminimal couplings with spin in the point 
particle action. Due to parity invariance the tensors, which contain the 
even-parity electric or the odd-parity magnetic curvature components, should 
be contracted with an even or odd number of the spin vector $S^{\mu}$ in 
eq.~\eqref{svec}, respectively, of an equal tensor rank. Yet, noting 
eq.~\eqref{ssquare}, one can equivalently use the square of the spin tensor 
$S^{\alpha}{}_{\mu} S^{\mu}{}_{\beta}$ instead of the tensor product of two 
spin vectors, since these differ by trace and 4-velocity terms, which vanish 
in the contraction with the traceless and orthogonal to $u^{\mu}$ curvature 
tensors. 

From reparametrization invariance and the definitions of the electric 
and magnetic components in eqs.~\eqref{Edef} and \eqref{Bdef}, we note that all 
these nonminimal couplings should be divided by the factor $u\equiv\sqrt{u^2}$.

Finally, as we noted for the minimal coupling part of the action, and at the 
end of section \ref{spinunGF}, these nonminimal couplings should be expressed 
in terms of the rotational gauge variables, using eq.~\eqref{Strans}.
 
\subsection{Nonminimal couplings to all orders in spin} \label{nonmcscq}

Based on the considerations from the previous section, we can actually 
write down the LO spin-induced nonminimal couplings in eq.~\eqref{Spp} to 
all orders in spin in the following:
\begin{align} \label{sinmc}
L_{\text{SI}}=&\sum_{n=1}^{\infty} \frac{\left(-1\right)^n}{\left(2n\right)!}
\frac{C_{ES^{2n}}}{m^{2n-1}} D_{\mu_{2n}}\cdots D_{\mu_3}
\frac{E_{\mu_1\mu_2}}{\sqrt{u^2}} S^{\mu_1}S^{\mu_2}\cdots 
S^{\mu_{2n-1}}S^{\mu_{2n}}\nn\\
&+\sum_{n=1}^{\infty} \frac{\left(-1\right)^n}{\left(2n+1\right)!}
\frac{C_{BS^{2n+1}}}{m^{2n}} 
D_{\mu_{2n+1}}\cdots D_{\mu_3}\frac{B_{\mu_1\mu_2}}{\sqrt{u^2}} 
S^{\mu_1}S^{\mu_2}\cdots 
S^{\mu_{2n-1}}S^{\mu_{2n}}S^{\mu_{2n+1}},
\end{align}
where in the first term of the first sum it is understood that the covariant 
derivatives do not exist. The mass exponent in the prefactor is actually set 
such that the Wilson coefficients are rendered dimensionless. The factorial 
in the prefactor is fixed from the symmetry of the spin-induced multipole. Also, 
the sign is alternating for each pair of spins that is added with a corresponding 
pair of derivatives, which can be seen from considering the definition of the multipole 
expansion in Fourier space, and then passing to coordinate space.   
In principle, the numerical factor is fixed such that the Wilson coefficients 
equal unity for the black hole case. It should be noted that the relation between 
all Wilson coefficients in eq.~\eqref{sinmc}, and the multipole moments used in numerical 
simulations \cite{Laarakkers:1997hb, Pappas:2012ns, Yagi:2014bxa}, should be worked
out using a formal matching procedure. It is clear from \cite{Pappas:2012ns},
that this can present subtleties, and is therefore left for future work.
However, the numerical values for the multipoles in
\cite{Laarakkers:1997hb, Pappas:2012ns, Yagi:2014bxa} are expected to lead to
good estimates for the Wilson coefficients through an ad hoc identification.

As we already noted these operators are naturally ordered by their PN 
relevance. For each multipole with N spins, we have N derivatives of the 
field, hence it is suppressed by the ratio $r_s^{(2N-N)}/r^N=v^{2N}$ with 
respect to the Newtonian point-mass term. Yet, in the magnetic odd-parity 
curvature component, which is coupled to the odd in spin multipoles, the LO
coupling must involve the gravito-magnetic vector of the NRG fields (see 
section \ref{tetradfield} below, and eq.~\eqref{nrgmetric} there, to recall the 
NRG fields), rather than the scalar, which appears as the LO coupling in the 
even-parity case \cite{Levi:2014gsa}. At LO the gravito-magnetic vector must 
be contracted with its LO mass coupling, which already carries one further 
power of $v$. Therefore, the even multipoles with $2n$ spins enter at the 
$(2n)$PN order, whereas the odd multipoles with $(2n+1)$ spins enter at the 
$(2n+1.5)$PN order for rapidly rotating compact objects.
 
Now, we are interested in particular in the spin-induced nonminimal couplings, 
which contribute up to the 4PN order for rapidly rotating compact objects, 
i.e.~in the quadrupole, octupole, and hexadecapole. The LO nonminimal coupling 
of the spin-induced quadrupole, which gives rise to the well-known LO 
spin-induced finite size effect \cite{Barker:1975ae,Poisson:1997ha}, was noted in \cite{Porto:2008jj}, though in a different form than what we derive here. 
The LO nonminimal couplings of the 
spin-induced octupole and hexadecapole were presented in \cite{Levi:2014gsa} 
only recently, where they were also obtained using the EFT formulation for spin, which we 
introduce here. Let us present then these LO nonminimal couplings explicitly. 

\paragraph{LO spin-squared.}
From eq.~\eqref{sinmc} we can read the LO nonminimal coupling for the 
spin-induced quadrupole as
\begin{align} \label{es2}
L_{ES^2} =& -\frac{C_{ES^2}}{2m} \frac{E_{\mu\nu}}{\sqrt{u^2}} 
S^{\mu} S^{\nu},
\end{align}
where the quadrupolar deformation constant due to spin from 
\cite{Poisson:1997ha}, introduced in \cite{Porto:2008jj} as the Wilson 
coefficient $C_{ES^2}$, equals unity in the black hole case. For 1.4 solar 
mass neutron stars a numerical computation yields $C_{ES^2}\simeq 4-8$, 
depending on the equation of state \cite{Laarakkers:1997hb,Pappas:2012ns}. 
This nonminimal coupling contributes as of the 2PN order for rapidly 
rotating compact objects. 
 
\paragraph{LO cubic in spin.}
From eq.~\eqref{sinmc} we read the LO nonminimal coupling for the spin-induced 
octupole as
\begin{align} \label{bs3}
L_{BS^3}=&-\frac{C_{BS^3}}{6m^2}\frac{D_\lambda B_{\mu\nu}}
{\sqrt{u^2}}S^{\mu} S^{\nu} S^{\lambda}.
\end{align}
This nonminimal coupling was introduced in \cite{Levi:2014gsa}, and also 
recently confirmed in an equivalent form in \cite{Marsat:2014xea}. The Wilson 
coefficient $C_{BS^3}$ 
introduced in \cite{Levi:2014gsa}, which encodes the octupolar deformation 
due to spin, equals unity in the black hole case, where its value for neutron 
stars was obtained in \cite{Pappas:2012ns, Yagi:2014bxa}.
This nonminimal coupling contributes as of the 3.5PN 
order for rapidly rotating compact objects. 

\paragraph{LO quartic in spin.}
Finally, from eq.~\eqref{sinmc} we also read the LO nonminimal coupling for the 
spin-induced hexadecapole as 
\begin{align} \label{es4}
L_{ES^4}=&\frac{C_{ES^4}}{24m^3} \frac{D_\lambda D_\kappa 
E_{\mu\nu}}{\sqrt{u^2}} S^{\mu} S^{\nu} S^{\lambda} S^{\kappa}.
\end{align}
This nonminimal coupling was also introduced in \cite{Levi:2014gsa} with the 
Wilson coefficient $C_{ES^4}$, which encodes the hexadecapolar deformation 
due to spin. Also here this Wilson coefficient equals unity in the black hole 
case, and its value for neutron stars was obtained in \cite{Yagi:2014bxa}. 
This nonminimal coupling contributes as of the 4PN order for rapidly rotating 
compact objects.

\section{Integrating out the orbital scale}\label{integrateoutorbital}

In order to obtain an EFT of radiation, including the conservative sector, we 
should proceed now to integrate out the field modes below the orbital scale. 
First, we recall that 
\begin{equation} 
\eta_{AB} \hat{e}^{A\mu} \hat{e}^{B\nu} = g^{\mu\nu},
\end{equation}  
that is the worldline tetrad contains field DOFs, in addition to 
the worldline rotational DOFs. In order to integrate out the 
field DOFs, we should disentangle them from the rotational DOFs. For that, we 
consider a tetrad field, $\tilde{e}_{a\mu}$, such that
\begin{equation}\label{emetric}
\eta_{ab} \tilde{e}^a{}_{\mu} \tilde{e}^b{}_{\nu} = g_{\mu\nu},
\end{equation}   
and we have that
\begin{equation}
\hat{e}_A{}^{\mu} = \hat{\Lambda}_A{}^b \tilde{e}_b{}^{\mu}, 
\end{equation}
where
\begin{equation}\label{LorCondition}
\eta^{AB} \hat{\Lambda}_A{}^a \hat{\Lambda}_B{}^b = \eta^{ab}.
\end{equation}
Then the rotational DOFs are contained in $\hat{\Lambda}_A{}^b$.
Notice that $\hat{e}_A{}^{\mu}$ and $\hat{\Lambda}_A{}^a$ are only defined on 
the worldline, whereas $\tilde{e}_a{}^{\mu}$ is a field over spacetime. Actually here 
$\tilde{e}_a{}^{\mu}$ is the fundamental field, unlike the nonspinning case, which can 
be formulated only in terms of the metric. An important consequence of this change 
in the representative DOFs of the field is an additional gauge freedom due to 
the internal Lorentz invariance of the local tetrad. This will be discussed in 
section \ref{tetradfield} below.

Let us then go on to disentangle the field from the worldline DOFs in the 
action. Then the minimal coupling term from eq.~\eqref{mcTrans} can be 
rewritten as \cite{Levi:2010zu} 
\begin{align} \label{Sfieldsplit}
\frac{1}{2} \hat{S}_{\mu\nu} \hat{\Omega}^{\mu\nu}
&= \frac{1}{2} \hat{S}_{ab} \hat{\Omega}^{ab}_{\text{flat}}
         + \frac{1}{2} \hat{S}_{ab} \omega_{\mu}{}^{ab} u^{\mu},      
\end{align}
where we have used the Ricci rotation coefficients, defined in 
eq.~\eqref{Riccirotation}, $\hat{\Omega}^{ab}_{\text{flat}} 
= \hat{\Lambda}^{Aa} \frac{d \hat{\Lambda}_A{}^b}{d \sigma}$ is the locally 
flat spacetime angular velocity tensor, and 
$\hat{S}_{ab}=\tilde{e}^{\mu}_{a}\tilde{e}^{\nu}_{b}\hat{S}_{\mu\nu}$ is the 
spin projected to the local frame. It should be stressed that one should 
also switch to the spin and angular velocity in the local frame as the 
fundamental variational variables. Note also that in this form the canonical 
SO(1,3) form already emerges, as the first term on the right hand side of 
eq.~\eqref{Sfieldsplit} is the kinematic term, which represents the Poisson 
brackets, and the Legendre transform in the angular velocity 
$\hat{\Omega}^{ab}_{\text{flat}}$ is already automatically performed.

Yet the separation of the field from the particle worldline DOFs is not 
complete. The gauge of the worldline temporal Lorentz matrix, 
$\hat{\Lambda}_{[0]}{}^a=w^a=\tilde{e}^a{}_\mu w^\mu$, may contain further 
field dependence, and the temporal components of the local spin also contain in 
general further field dependence \cite{Levi:2014sba}. We recall that 
following \cite{Yee:1993ya}, in the application of a Routhian approach for spin in \cite{Porto:2008tb, Porto:2008jj} the SSC are applied 
at the level of the EOM, similar to traditional methods working at the 
level of the EOM, see e.g.~\cite{Faye:2006gx, Marsat:2012fn, Bohe:2012mr}. 
Therefore, similarly both require further computation of the metric or 
tetrad field in order to extract the physical EOM \cite{Levi:2014sba}. This 
is in conflict with the definition of the EFT in eq.~\eqref{orbitintegrateout}, 
since that means that the field has actually not been completely integrated 
out, and indeed an EFT, i.e.~an effective action at the orbital scale has 
not been obtained. As was put forward and implemented in \cite{Levi:2008nh}, 
the field would be completely disentangled from the worldline DOFs once a gauge 
for the worldline rotational variables is fixed at the level of the Feynman rules. 
It should be noted that also in the ADM Hamiltonian formalism the constraints for 
the spin and for its conjugate DOFs are applied prior to the obtainment of the potential, 
e.g.~in \cite{Damour:2007nc, Steinhoff:2009ei}.
The gauge fixing of the rotational variables will be further illustrated in section \ref{spinGF} below.  
 
\subsection{Tetrad field gauge} \label{tetradfield}

We recall from eqs.~\eqref{orbitfreq} and \eqref{orbitmomentum} that the field 
modes at the orbital scale are instantaneous at leading order. Therefore, a 
Kaluza-Klein like reduction over the time dimension at this stage makes a very 
sensible decomposition of spacetime in the nonrelativistic limit 
\cite{Kol:2007bc, Kol:2010ze}. This parametrization of the metric, given by   
\be \label{nrgmetric}
d s^2 = g_{\mu\nu} dx^{\mu} dx^{\nu} \equiv 
e^{2 \phi}(dt - A_idx^i)^2 - e^{-2 \phi}\gamma_{ij}dx^i dx^j, 
\ee
defines the nonrelativistic gravitational (NRG) fields $\phi$, $A_i$, and 
$\gamma_{ij}\equiv\delta_{ij}+\sigma_{ij}$. It should be noted that an exponential 
parametrization of the metric coefficients was already introduced in 
\cite{Damour:1990pi}. Indeed, the NRG fields proved to 
be advantageous for PN applications both in the non-spinning 
\cite{Kol:2007bc, Kol:2010ze}, and spinning cases 
\cite{Levi:2008nh,Levi:2010zu,Levi:2014gsa}. 
In addition to providing physical insight and exhibiting a useful hierarchy in the 
worldline couplings, the NRG fields simplify the propagators, the interaction 
vertices, and their extraction. We will also see here in section \ref{nloss}, 
that the NRG fields reduce the number of Feynman diagrams in topologies of 
higher loop order. In addition, the worldline mass couplings are simple and 
immediate to obtain. We would like to have these benefits for the worldline 
spin couplings as well, hence we should gauge the tetrad field in a 
sensible manner adapted to the NRG space+time decomposition of the metric. 

The time gauge of Schwinger \cite{Schwinger:1963re}, which is also used with the ADM space+time field decomposition, see e.g.~in \cite{Steinhoff:2009ei}, provides such a sensible gauge for a tetrad of NRG fields. Using the internal local Lorentz 
invariance of the tetrad field, the time gauge is defined by locking the time axes 
of the local coordinate systems to the  time axis of the general coordinate systems. 
We take $\tilde{e}_a{}^0$ to be a timelike vector in the local frame, and then 
it is possible to choose each local frame so that the spatial components of 
$\tilde{e}_a{}^0$ vanish. That is, in the time gauge we take
\be \label{tgauge}
\tilde{e}_{(i)}{}^0(x)=0.
\ee 
Then, we have from eqs.~\eqref{emetric}, \eqref{nrgmetric}, and 
\eqref{tgauge}, and again using the local Lorentz symmetry, that all the 
components of the tetrad can be chosen as
\begin{equation} \label{NRGtetrad}
\tilde{e}^a{}_{\mu} = \left( \begin{array}{cc}
e^{\phi} & - e^{\phi} A_i \\
0 & e^{-\phi} \sqrt{\gamma}_{ij}
\end{array} \right),
\end{equation}
where $\sqrt{\gamma}_{ij}$ is the symmetric square root of $\gamma_{ij}$, 
for which we should solve. Indeed, using the 3+3 degrees of gauge freedom of 
the internal Lorentz symmetry, we are left with the 10 DOFs of the metric, 
out of the 16 DOFs of a general tetrad. Hence, we have a closed form tetrad 
for the NRG fields apart from $\sqrt{\gamma}_{ij}$, which is conveniently 
defined, with some trivial components 
$\tilde{e}^{a}{}_0 = e^{\phi} (1, \vec{0})$. 

\subsection{Fixing the gauge of the rotational variables} \label{spinGF}

In a classical setting one can fix the rotational gauge by simply inserting 
the rotational constraints in the action before integrating out the field. 
It should be stressed that if the rotational gauge is not fixed prior to 
integrating out the field, the orbital field still appears at the level of 
the EOM, where the SSC should be applied, indicating that the EFT 
computation is incomplete \cite{Levi:2014sba}, and see also section 
\ref{spineom} below. 

We recall from eq.~\eqref{Sfieldsplit}, that we now use the Lorentz matrices, 
connecting the worldline and local frames, $\hat{\Lambda}_A{}^a$, 
and the spin $\hat{S}_{ab}$ projected to the local frame, as our rotational 
variables. These still contain gauge freedom, that should be fixed by the gauge 
conditions applied in the local frame, given in the form 
\begin{align} 
\hat{\Lambda}_{[0]a} &= w_{a},\label{gengaugelocal}\\
\hat{S}^{ab} \left( p_{b} + \sqrt{p^2} \hat{\Lambda}_{[0]b} \right) &= 0.
\label{genSSClocal}
\end{align}
For similar conditions in the flat spacetime case, see \cite{Steinhoff:2015ksa}.
As we already noted these gauge constraints should be implemented at the level 
of the action in order to ultimately arrive at an effective action without any 
remaining orbital scale field modes. We will also see now, that we cannot 
obtain an effective action formulated with the worldline spin as a further DOF, 
if we gauge fix the spin without gauge fixing its conjugate DOFs.

We go on now to consider the three sensible options to fix this gauge: the 
covariant gauge, the canonical gauge, and the gauge of no mass dipole, which 
we formulate here below. The canonical gauge, which is associated with 
canonical variables, is of particular interest due to the advantages of such 
variables for the obtainment of the spin EOM, and of a corresponding 
Hamiltonian. We will see that in either case, we will ultimately have to 
switch to the rotational variables in the canonical gauge in order to 
completely integrate out the field, and get an effective action. Intuitively, 
this is the consequence of having our point particle action in eq.~\eqref{Spp}, 
formulated in terms of the spin in addition to its conjugate as independent 
variables, as we noted for the form of eq.~\eqref{Sfieldsplit}.    

\paragraph{Covariant gauge.}

We can choose in eq.~\eqref{gengaugelocal} 
\be
\hat{\Lambda}_{[0]a} =\frac{p_a}{\sqrt{p^2}}. 
\ee
In this case we are back with Tulczyjew's covariant SSC, 
and the original Lorentz matrices and spin variables, 
such that it holds for the Lorentz matrices that
\begin{equation} \label{covgauge}
\Lambda_{[0]}{}^a =\frac{p^{a}}{\sqrt{p^2}}\Leftrightarrow\Lambda_{[i]a}p^{a}=0.
\end{equation}

Yet, we note that in this case the matrix $\Lambda_{[i]}{}^{(j)}$ is not an SO(3) 
rotation matrix. Instead, from eqs.~\eqref{LorCondition} and \eqref{covgauge}, we 
get that
\begin{equation}
\Lambda_{[k]}{}^{(i)}\Lambda^{[k](j)}=\delta^{ij}+\frac{p^{(i)} p^{(j)}}{p^2}.
\end{equation}
Hence the Lorentz matrices depend on the field through 
$p^{a} = \tilde{e}^{a}{}_{\mu} p^{\mu}$.
This is not desirable for an EFT approach, as the field DOFs are not separated 
from the particle DOFs, which would hinder integrating out the field. Moreover, 
$\Lambda_{[i]}{}^{(j)}$ is mixing rotational and linear motion DOFs. We can 
already note that the only way out of this predicament is to redefine 
$\Lambda_{A}{}^{a}$ by boosting to the local rest frame. This 
is what we shall do now, and what is done in the canonical gauge, which 
we discuss next. 

Thus, let us stick here with the covariant gauge for the spin variable. On 
the upside, we note that the extra term from minimal coupling in 
eq.~\eqref{mcextra} drops, since the SSC is implemented in the action. Then, 
let us evaluate $\frac{1}{2} S_{ab} \Omega^{ab}_{\text{flat}}$ from 
eq.~\eqref{Sfieldsplit}. Using eq.~\eqref{SmcTrans} for the locally flat 
frames, with $w^a = \delta_0^a$ and $q^a = p^a / \sqrt{p^2}$, and 
implementing the covariant SSC in the action, we then obtain
\begin{align}\label{Smcflatold}
\frac{1}{2} S_{ab} \Omega^{ab}_{\text{flat}} =&
	\frac{1}{2} \left( S_{(i)(j)} - \frac{S_{(i)(k)}p^{(k)}p^{(j)}}
	{p_{(0)}\left(\sqrt{p^2}+p_{(0)}\right)}
		                            + \frac{S_{(j)(k)}p^{(k)}p^{(i)}}
  {p_{(0)}\left(\sqrt{p^2}+p_{(0)}\right)}\right) 
		\hat{\Omega}^{(i)(j)}_{\text{flat}}\nn\\
	  &+ \frac{S_{(i)(j)}p^{(j)}}{p_{(0)}\left(\sqrt{p^2}+p_{(0)}\right)} 
	    \frac{d p^{(i)}}{d\sigma},
\end{align}
where the temporal terms of $\Omega^{ab}_{\text{flat}}$ were dropped due to 
taking $w^a = \delta_0^a$, and 
\be
\hat{\Omega}^{(i)(j)}_{\text{flat}} 
= \hat{\Lambda}^{[k](i)} \frac{d \hat{\Lambda}_{[k]}{}^{(j)}}{d \sigma}.
\ee 

This is precisely doing the transformation of the Lorentz matrix in eq.~(4.5), 
using eqs.~(4.2), (4.3) from \cite{Levi:2014sba}, where the matrix 
$\hat{\Lambda}_{[i]}{}^{(j)}$ is now an SO(3) rotation matrix. Indeed, if we 
also make the spin transformation from eq.~(4.6) of \cite{Levi:2014sba} in 
eq.~\eqref{Smcflatold}, we get agreement with eq.~(4.7) there. Let us note 
then that also in eq.~\eqref{Smcflatold} here, we get an additional term with a 
derivative of the momentum instead of the extra term from minimal coupling in 
eq.~\eqref{mcextra}, that dropped due to the covariant SSC implemented. It is 
also already clear that $\hat{\Omega}^{(i)(j)}_{\text{flat}}$ will appear in 
the spin vertices, and therefore also in the final potential. Therefore, it is 
clear that the spin variable should indeed be redefined as in 
\cite{Levi:2014sba} in order to simplify the Feynman rules, and to easily 
obtain the EOM of spin, see section \ref{spineom}. 
This ultimately amounts to going to the canonical gauge. 
 
\paragraph{Canonical gauge.}

Let us choose in eq.~\eqref{gengaugelocal} the canonical gauge, that is  
\begin{equation} \label{pnwgauge}
\hat{\Lambda}_{[0]}{}^a= \delta_0^a 
\Leftrightarrow 
\hat{\Lambda}_A{}^{(0)} = \delta_A^0.
\end{equation}
Essentially, this means that we have boosted the Lorentz matrices $\Lambda_{Aa}$ 
to the local rest frame, such that they became rotation matrices, decoupled from the 
linear motion DOFs, and also from the field DOFs. From eqs.~\eqref{pnwgauge} and 
\eqref{TulczyjewSTrans} in the local frame we also get a new spin variable 
$\hat{S}_{ab}$, which from eqs.~\eqref{pnwgauge} and \eqref{genSSClocal} 
satisfies the canonical SSC, given in the local frame by
\be \label{pnwssc}
\hat{S}^{ab} \left( p_{b} + \sqrt{p^2} \delta_{0b} \right) = 0.
\ee
This is a generalization of the Pryce-Newton-Wigner SSC 
\cite{Pryce:1948pf,Newton:1949cq} for curved spacetime. 
It should be noted that a similar SSC was suggested in eq.~(4.7) of 
\cite{Barausse:2009aa} for a canonical formalism of a test particle at linear 
order in the particle spin. Note that their eq.~(4.7), formulated in terms of the 
local tetrad field, is not to be confused with eq.~\eqref{genSSC} here, which is a 
generic SSC, formulated in terms of the worldline tetrad, where the gauge of the 
worldline tetrad is not fixed. Further, in \cite{Barausse:2009aa} they choose the 
conjugate gauge of the worldline Lorentz matrices as the covariant one. Here the 
choice of gauge for the worldline tetrad or Lorentz matrices fixes also the spin 
gauge, namely the corresponding SSC.  

Then again similar to eq.~\eqref{Smcflatold} the temporal components of the 
Lorentz matrices simply drop, and we get
\be \label{mclocalspatial}
\frac{1}{2} \hat{S}_{ab} \hat{\Omega}^{ab}_{\text{flat}}
        = \frac{1}{2} \hat{S}_{(i)(j)} \hat{\Omega}^{(i)(j)}_{\text{flat}}. 
\ee
This means that we have now the familiar kinematics of a three-dimensional top.  
The canonical gauge is then useful to disentangle the particle and field 
DOFs, and is therefore optimal for a formulation of the EFT in terms of the 
worldline spin, since as discussed for the covariant gauge above, ultimately we 
have to switch to the rotational variables in the canonical gauge. Hence, we 
shall fix the spin gauge to the canonical one.

\paragraph{No mass dipole gauge.}

As a further illustration for a non-covariant spin gauge, we can choose in 
eq.~\eqref{gengaugelocal} 
\be \label{nomdipole}
\hat{\Lambda}_{[0]}{}^a=\frac{2p_0\delta_0^a-p^a}{\sqrt{p^2}}.
\ee
In this case we are boosting to the local frame, where the mass dipole vanishes, 
and from eqs.~\eqref{nomdipole} and \eqref{genSSClocal} we have 
\be
\hat{S}_{a0}=0.
\ee 
This is the SSC known from \cite{Corinaldesi:1951pb}.

\paragraph{Shift of the position coordinate.}

Once a specific non-covariant gauge is fixed for the rotational DOFs, the 
`center' of the spinning object, with respect to which the multipoles are 
considered, will no longer coincide with the spatial origin of the worldline 
tetrad. 
Hence one should eventually make a 
shift in the position coordinate from the worldline coordinate to the 
position coordinate of the `center'. This shift of position, would actually 
be enforced by the reduction of acceleration terms, originating from 
the extra minimal coupling term in eq.~\eqref{mcextra}. It is sensible then 
to make this shift of position only after the EFT computation is through, so 
that there is no ambiguity in shifting from the worldline coordinate, and in 
the implementation of the rotational gauge fixing at the level of the point 
particle action. We shall see in section \ref{implementation}, that 
performing the linear shift of position just corresponds to the insertion of 
lower order EOM for the position. Yet, it should be noted that the quadratic 
shift of individual positions would contribute to the spin-squared interaction 
as of the NLO.
   
As we noted in section \ref{nmcs} the nonminimal couplings in the action are 
naturally expressed in terms of the projected rotational variables, which 
we fix to the covariant gauge. Let us then work out the components of the spin, 
$S_{ab}$, in terms of the spin variable in the canonical gauge. From the 
canonical SSC in eq.~\eqref{pnwssc} we have 
\begin{equation} \label{pnws0i}
\hat{S}_{a(0)} = -\frac{\hat{S}_{a(i)} u^{(i)}}{u+u^{(0)}}
               = -\hat{S}_{ab} \frac{u^b}{u}, 
\end{equation}
and using this in eq.~\eqref{Strans}, we find
\begin{align}
S_{(i)(j)} &= \hat{S}_{(i)(j)}
        -\hat{S}_{(i)(k)}\frac{u_{(j)}u^{(k)}}{u\left(u+u^{(0)}\right)}
        +\hat{S}_{(j)(k)}\frac{u_{(i)}u^{(k)}}{u\left(u+u^{(0)}\right)}, \\
S_{(i)(0)} &= - \frac{\hat{S}_{(i)(j)} u^{(j)}}{u}. 
\end{align}
Note that here all velocities are the local tetrad projected ones, not the 
coordinate velocities.
Naturally, one would obtain similar relations upon using the covariant gauge 
first to eliminate the $S^{0i}$ components, and then transforming to the 
canonical gauge variables at the 3-dimensional level as in 
\cite{Levi:2014sba}. 
Similarly, using eq.~\eqref{pnwgauge} in eq.~\eqref{etrans}, leads to eq.~(4.5) of 
\cite{Levi:2014sba}, which we write here as
\begin{align}
\Lambda^{[i](j)} &= \hat{\Lambda}^{[i](k)} \left( \delta^{(k)(j)}
        + \frac{u^{(k)} u^{(j)}}{u\left(u+u^{(0)}\right)} \right).
\end{align}

Let us also work out explicitly the minimal coupling term in terms of the 
spatial components of the local spin variable in the canonical gauge. From 
eqs.~\eqref{mcTrans}, \eqref{mcextra}, and \eqref{Sfieldsplit}, we already have 
\begin{align} \label{mcsimplify}
\frac{1}{2} S_{\mu\nu} \Omega^{\mu\nu} &= 
     \frac{1}{2} \hat{S}_{ab} \hat{\Omega}^{ab}_{\text{flat}}
   + \frac{1}{2} \hat{S}_{ab} \omega_{\mu}{}^{ab} u^{\mu} 
   + \frac{\hat{S}_{ab} u^b}{u^2} \frac{D u^a}{D \sigma}, 
\end{align}
and note that
\be \label{extramcsimplify} 
\frac{D u^a}{D \sigma}= \dot{u}^a + \omega_{\mu}{}^{ca} u_c u^{\mu}. 
\ee
Next we use eqs.~\eqref{mclocalspatial} and \eqref{pnws0i} for the canonical 
gauge, and get from eqs.~\eqref{mcsimplify} 
and \eqref{extramcsimplify} that
\begin{align} \label{mcsimplified}
\frac{1}{2} S_{\mu\nu} \Omega^{\mu\nu} &=
	  \frac{1}{2} \hat{S}_{ab} \hat{\Omega}^{ab}_{\text{flat}}
	- \frac{\hat{S}_{ab} u^a \dot{u}^b}{u^2}
  + \frac{1}{2} \omega_{\mu}{}^{ab} u^{\mu} \left(
        	\hat{S}_{ab} - 2 \hat{S}_{ac}\frac{u^c u_b}{u^2} \right) \nn\\
&= \frac{1}{2} \hat{S}_{(i)(j)} \hat{\Omega}^{(i)(j)}_{\text{flat}}
	- \frac{\hat{S}_{(i)(j)} u^{(i)} \dot{u}^{(j)}}{u\left(u+u^{(0)}\right)}\nn\\
&\quad    + \frac{1}{2} \omega_{\mu}{}^{(i)(j)} u^{\mu} \left(\hat{S}_{(i)(j)} 
   - 2 \frac{\hat{S}_{(i)(k)} u^{(k)}u_{(j)}}{u\left(u+u^{(0)}\right)} \right)
        	+ \omega_{\mu}{}^{(0)(i)} u^{\mu} \frac{\hat{S}_{(i)(j)} u^{(j)}}{u}.
\end{align}
Also here for the minimal coupling term it naturally holds, that the same 
relation is obtained upon first eliminating the $S^{0i}$ components with the 
covariant gauge, and then using the 3-dimensional transformations to the 
canonical gauge variables as in \cite{Levi:2014sba}.   

\subsection{Feynman rules} \label{Feynmanrules}

From the previous sections \ref{nmcs}, \ref{tetradfield}, and \ref{spinGF}, 
we can derive all the Feynman rules required up to quadratic level in the spin 
to NLO. For the nonminimal couplings, which are cubic and quartic in the spin from 
section \ref{nmcs}, and the consequent Feynman rules to LO, we refer to 
\cite{Levi:2014gsa}, where the cubic and quartic in spin sectors were fully 
obtained for generic compact binaries. For the Feynman rules, concerning the 
purely gravitational action in harmonic gauge, and the point-mass nonspinning 
action worldline couplings from eq.~\eqref{Spp}, we refer to section II of 
\cite{Levi:2010zu}. Similar to \cite{Levi:2010zu,Levi:2014gsa} we use here the 
NRG fields \cite{Kol:2007bc, Kol:2010ze}, and the related tetrad field in the 
time gauge introduced in section \ref{tetradfield}. Similarly, for the 
worldline affine parameter we choose the coordinate time $t=y^0$, 
i.e.~$\sigma=t$, such that we have $u^0=1$, $u^i=dy^i/dt\equiv v^i$. Here we 
give the Feynman rules for the worldline spin couplings, since only 
these are modified with respect to previous works, such that they are given 
here explicitly in terms of the spatial components of the local spin variable 
in the canonical gauge. Hence from now on we drop the hat notation on the rotational variables, and the round brackets on their indices, and in addition all indices are Euclidean. 

From eq.~\eqref{mcsimplified} we see that we have contributions from kinematic 
terms involving spin without field coupling. To the order we are considering, these are given by
\be \label{frskin}
L_{\text{kin}}=-\vec{S}\cdot\vec{\Omega}
         - \frac{1}{2} \left(1+\frac{3}{4}v^2\right) \epsilon_{ijk}S_k v^j a^i,
\ee
where $S_{ij}=\epsilon_{ijk}S_k$, $\Omega_{ij}=\epsilon_{ijk}\Omega_k$, 
and $a^i\equiv\dot{v}^i$. 
Then, for the one-graviton couplings to the worldline spin the Feynman rules 
required in this work to NLO are 
\begin{align}
\label{eq:sA}  \parbox{12mm}{\includegraphics[scale=0.6]{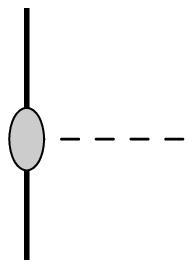}}
 & = \int dt \,\, \epsilon_{ijk}S_k\left(\frac{1}{2}\partial_iA_j + \frac{3}{4} v^iv^l 
 \left(\partial_lA_j-\partial_jA_l\right)+v^i\partial_tA_j\right), \\ 
\label{eq:sphi}   \parbox{12mm}{\includegraphics[scale=0.6]{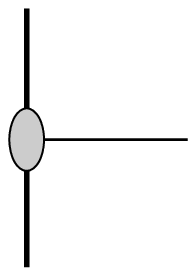}}
 & = \int dt \,\,\epsilon_{ijk}S_k v^i\left(2\partial_j\phi + v^2 \partial_j\phi 
 - 2a^j\phi\right), \\  
\label{eq:ssigma}   \parbox{12mm}{\includegraphics[scale=0.6]{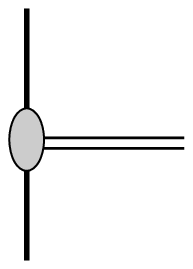}}
 & = \int dt \,\,\frac{1}{2}\epsilon_{ijk}S_k v^l\partial_i\sigma_{jl},
\end{align}
where the oval (gray) blob represents the spin on the worldline. For the 
two-graviton couplings to the worldline spin, the Feynman rule required here 
to NLO is: 
\begin{align}
\label{eq:sphiA}  \parbox{12mm}{\includegraphics[scale=0.6]{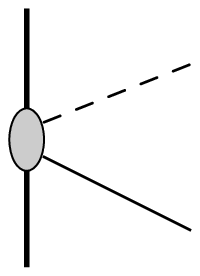}}
 & = \int dt \,\,2\epsilon_{ijk}S_k\,\phi\,\partial_iA_j. 
\end{align}
Note that the two-scalar coupling to the worldline spin vanishes in our gauge, 
which is a desirable feature, reducing the number of total diagrams.

From the $ES^2$ nonminimal coupling in eq.~\eqref{es2} the Feynman rules for 
the one-graviton couplings to the worldline spin-squared are given by
\begin{align}
\label{eq:sqphi} \parbox{12mm}{\includegraphics[scale=0.6]{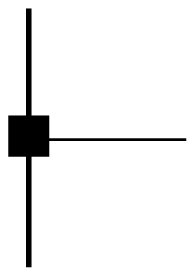}}
 & = \int dt \,\,\frac{C_{ES^2}}{2m}\left[S^{i}S^{j}\left(
 \partial_i\partial_j\phi \left(1+\frac{3}{2}v^2\right)
 -3\partial_j\partial_k\phi\,v^iv^k
 -2\partial_i\partial_t\phi\,v^j\right)\right.\nn\\
 &\qquad\qquad\qquad-S^2\left(
 \partial_i\partial_i\phi\left(1+\frac{3}{2}v^2\right)
 -\partial_i\partial_j\phi\, v^iv^j
 +2\partial_i\partial_t\phi\,v^i
 +2\partial_t^2\phi\right)\bigg],\\
\label{eq:sqA}  \parbox{12mm}{\includegraphics[scale=0.6]{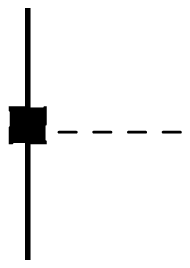}}
 & = \int dt \,\, -\frac{C_{ES^2}}{2m}\left[S^{i}S^{j}
 \left(\partial_i\partial_j A_kv^k-\partial_i\partial_k A_jv^k
 -\partial_i\partial_t A_j\right)\right.\nn\\
 &\qquad\qquad\qquad\qquad\left.-S^2\left(
 \partial_i\partial_i A_kv^k-
 \partial_i\partial_k A_iv^k
 -\partial_i\partial_t A_i\right)\right], 
\end{align}
where the square (black) box represents the $ES^2$ spin coupling on the worldline. 
Note that the last terms in the first and second lines of eq.~\eqref{eq:sqphi}, 
and the last four terms on eq.~\eqref{eq:sqA}, are missing from eqs.~(40), (39), 
respectively, in \cite{Porto:2008jj}. 
We should stress that time derivatives of spin should not be dropped before
all higher order time derivatives are treated rigorously in the resulting action.
Finally, for two-graviton couplings from the worldline $ES^2$ term the Feynman 
rule required for the NLO spin-squared sector is: 
\begin{align}
\label{eq:sqphi2}   \parbox{14mm}{\includegraphics[scale=0.6]{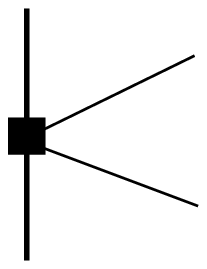}}
 & = \int dt\,\,\frac{C_{ES^2}}{2m}\left[ 
 3S^{i}S^{j}\left(\partial_i\phi \partial_j\phi 
 + \phi\,\partial_i\partial_j\phi\right) 
 -S^2 \left(\left(\partial_k\phi\right)^2 
 +3\phi\,\partial_i\partial_i\phi\right)\right].
\end{align}

\subsection{EOM of the spin} \label{spineom}

Using the Feynman rules detailed in the previous section to evaluate the 
relevant Feynman diagrams, a spin dependent effective action is obtained of 
the form:  
\be
S_{\text{eff(spin)}}=\int dt\left[-\frac{1}{2}\sum_{I=1}^2 S_{Iij}
\Omega_I^{ij}-V\left(\vec{x}_I,\dot{\vec{x}}_I, \ddot{\vec{x}}_I, 
\dots, S_{Iij}, \dot{S}_{Iij}, \dots \right)\right],
\ee 
where here we express the result again in terms of the spin tensor, using  $S_i=\frac{1}{2}\epsilon_{ijk}S_{jk}$.
As explained already in sections 3.1, 5.2 of \cite{Levi:2014sba}, similar to 
the EOM of the position, the EOM of the spin should be obtained from a 
variation of the action, in terms of which the EFT approach is naturally 
formulated. It should be underlined that one should make an independent 
variation of this action with respect to the spin, and to its conjugate 
variables, the rotation matrices.
Then the following simple form for the EOM 
of the spin is obtained
\begin{align}\label{Seom}
\dot{S}_I^{ij}=&
- 4 S_I^{k[i} \delta^{j]l}\frac{\delta\int{dt\,V}}{\delta S_I^{kl}}
=- 4 S_I^{k[i} \delta^{j]l} \left[\frac{\partial V}
{\partial S_I^{kl}} - \frac{d}{dt} \frac{\partial V}{\partial \dot{S}_I^{kl}} 
+ \dots\right],
\end{align}
see eq.~(5.9) in \cite{Levi:2014sba}.

It should be stressed that the spin EOM in eq.~\eqref{Seom} are free of the unphysical 
spin DOFs $S^{0i}$, and that this is essential in an EFT which removes the orbital 
scale field.
However, if one disregards this basic requirement of an EFT, 
one might also leave the $S^{0i}$ components of the spin tensor 
as independent DOFs till after the obtainment of the EOM, as advocated in \cite{Porto:2008tb}.
In that case the spin dependent action obtained after the evaluation of the 
relevant Feynman diagrams is of the form \cite{Levi:2014sba}:
\be
S_{\text{(spin)}}=\int dt \left[ -\frac{1}{2}\sum_{I=1}^2S_{Iab}
\Omega_I^{ab}-V\left(\vec{x}_I,\dot{\vec{x}}_I, \ddot{\vec{x}}_I, \dots, 
S_{Iab}, \dot{S}_{Iab}, \dots \right)\right]. 
\ee 
If we follow the Routhian approach in \cite{Porto:2008tb}, then we should 
derive the EOM of the spin in terms of the Poisson brackets of the 
so$(1,3)$ spin algebra. Since as of NLO we have higher order time 
derivatives of the spin in the potentials, which actually contribute in the 
potentials as of NNLO, it is in fact improper to derive the EOM of the spin 
using Poisson brackets \cite{Levi:2014sba}. 

Again, we can instead make an independent variation of this action with 
respect to the spin, and to its conjugate Lorentz matrices, to obtain 
the following EOM of the 4-dimensional spin tensor \cite{Levi:2014sba}: 
\begin{align}\label{S4deom} 
\dot{S}_I^{ab}=&\, 
4 S_I^{c[a} \eta^{b]d}\frac{\delta\int{dt\,V}}{\delta S_I^{cd}} 
= 4 S_I^{c[a} \eta^{b]d} \left[ \frac{\partial V}{\partial S_I^{cd}} 
- \frac{d}{dt} \frac{\partial V}{\partial \dot{S}_I^{cd}} + \dots \right]. 
\end{align}
However and more importantly, in that case after the obtainment of the EOM a 
further EFT computation of the metric at the orbital scale would be required 
in order to eliminate the $S^{0i}$ components, and to extract the physical 
EOM, as was demonstrated in section 3 of \cite{Levi:2014sba}, since the 
$S^{0i}$ components contain orbital scale field DOFs. 
This situation is actually similar to traditional methods, e.g.~in 
\cite{Faye:2006gx,Marsat:2012fn,Bohe:2012mr}, where the unconstrained EOM 
are derived in harmonic coordinates, and then the SSC are inserted, using the 
metric, which is explicitly computed for the derivation of the unconstrained EOM.
This implies that the 
EFT computation is in fact incomplete when a so(1,3) potential is presented.  
In the current formulation of an EFT for spin we avoid this undesirable outcome, 
and directly obtain the physical EOM in the form of eq.~\eqref{Seom}.

\section{Spin potentials to NLO} \label{implementation}

In this section we begin to implement the EFT for gravitating spinning objects, 
which we formulated in the previous sections. We start with the computation of 
all spin dependent potentials up to quadratic level in the spin to NLO, that is 
to the 3PN order for rapidly rotating compact objects. We will use the Feynman 
rules from section \ref{Feynmanrules}, with the NRG fields, and the canonical 
spin gauge. In particular, we also detail here the NLO spin-squared sector, 
where the NRG fields also turn out to be advantageous, and to remove one-loop 
topology diagrams, similar to the non-spinning point mass case.

We start by evaluating the Feynman diagrams, as well as the kinematic 
contributions from eq.~\eqref{frskin}, for each of the sectors. We obtain 
potentials, which contain only physical DOF, and higher order time 
derivatives of the variables of the particles, just like the non-spinning 
potentials. From this point these potentials are handled in the standard 
manner in a PN scheme.
The EOM of the position, and of the spin, can then be directly obtained 
via a proper variation of the action, see section \ref{spineom}. 
The higher order time derivatives should be removed at the level of the EOM 
then, so that the EOM are well-defined. 
Equivalently, here we first eliminate higher order time derivatives via 
redefinitions of the position and spin variables at the level of the action, 
see section 5 in \cite{Levi:2014sba}. Again, the EOM of the position and 
spin can be obtained via a proper variation of the action. Further, the 
Hamiltonian is obtained by a straightforward Legendre transform.      

\subsection{LO spin sectors} \label{loso}

\subsubsection{LO spin-orbit sector}

The LO spin-orbit sector has been worked out in similar terms in section 4 of 
\cite{Levi:2010zu} with the final action in eq.~(72) there. We have the same 
two Feynman diagrams, shown in figure 1, such that from the diagrams we get the 
total value presented in eq.~(71) of \cite{Levi:2010zu}. 
\begin{figure}[t]
\begin{center}
\includegraphics[scale=0.5]{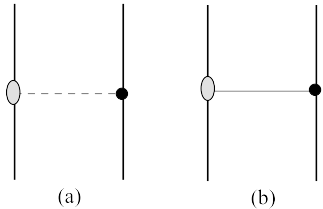}
\caption{LO spin-orbit Feynman diagrams. These diagrams should be included 
together with their mirror images. Note that the value of diagram (a) does not 
change with respect to \cite{Levi:2010zu}.} 
\end{center}
\label{fig:solo}
\end{figure}
Note that the value of diagram (a) does not depend on the spin gauge, and can 
be found in eq.~(65) of \cite{Levi:2010zu}. The value of diagram (b), which 
depends on the spin gauge, is given by 
\begin{align}
\text{Fig.~1(b)}=&2\frac{Gm_2}{r^2}\vec{S}_1\cdot\vec{v}_1\times\vec{n} + 
1\leftrightarrow2,
\end{align}
where $\vec{r}\equiv\vec{y}_1-\vec{y}_2$, $r\equiv\sqrt{{\vec{r}}^{2}}$, and 
$\vec{n}\equiv\vec{r}/r$.
In addition, we have a kinematic contribution in eq.~\eqref{frskin} from the 
extra minimal coupling term as noted in eq.~(72) of \cite{Levi:2010zu}, which 
equals
\be
L_{\text{kin}}=\frac{1}{2}\vec{S}_1\cdot\vec{v}_1\times\vec{a}_1
+1\leftrightarrow2,
\ee 
and is acceleration dependent.

\paragraph{The potential and Hamiltonian.}

All in all, as in eq.~(72) of \cite{Levi:2010zu}, we obtain the potential
\begin{align}\label{loso}
V_{\text{SO}}^{\text{LO}}=&-2\frac{Gm_2}{r^2}\vec{S}_1\cdot\left[\vec{v}_1
\times\vec{n}-\vec{v}_2\times\vec{n}\right] - \frac{1}{2}\vec{S}_1\cdot
\vec{v}_1\times\vec{a}_1 +1\leftrightarrow2. 
\end{align}
The EOM can be obtained via a variation of the action. 
We go on to perform a shift of the positions, $\Delta \vec{y}_I$, according to
\begin{equation} \label{positionshift}
\vec{y}_1 \rightarrow \vec{y}_1 + \frac{1}{2 m_1}\vec{S}_1\times\vec{v}_1,
\end{equation}
and similarly for particle 2 with $1\leftrightarrow2$, corresponding to the 
shift in eq.~\eqref{covariantshift}. The contribution to the action, which is 
linear in the shifts, removes the acceleration terms, and is equivalent to 
substituting in the Newtonian EOM of the position. It reads
\begin{align}\label{lososhift}
\Delta V_{\text{SO}}^{\text{LO}}(\Delta \vec{y}_{I})=& 
	 \frac{1}{2} \vec{S}_1\cdot\vec{v}_1\times\vec{a}_1 
	 + \frac{G m_2}{2r^2} \vec{S}_1\cdot\vec{v}_1\times\vec{n} 
	 + 1 \leftrightarrow 2.
\end{align}
The potential in eqs.~\eqref{loso} and \eqref{lososhift} was first derived in \cite{Tulczyjew:1959}.
Then, with a trivial Legendre transform we obtain the Hamiltonian 
\begin{align}
H_{\text{SO}}^{\text{LO}}=&-\frac{Gm_2}{r^2}\vec{S}_1\cdot\left[\frac{3}{2}
\frac{\vec{p}_1\times\vec{n}}{m_1}-2\frac{\vec{p}_2\times\vec{n}}{m_2}\right]
+1\leftrightarrow2. 
\end{align}
We note that at this point the EOM can be obtained from the Poisson brackets, 
where we have now that 
the position and momentum variables are canonical conjugate to each other, namely
\be
\{y_I^i,p_J^j\}= \delta^{ij}\delta_{IJ},
\ee
and that the spin variables also satisfy the canonical Poisson bracket relations, 
namely
\be
\{S_I^i,S_J^j\}=\epsilon^{ijk}S_I^k\delta_{IJ}.
\ee

Note that further terms linear in each of the shifts in eq.~\eqref{positionshift} 
contribute to all NLO spin sectors, and correspond to substituting in EOM 
from the non-spinning 1PN order, and LO SO sectors. They also contribute to 
the LO cubic in spin sector, corresponding to the insertion of EOM from the LO 
spin1-spin2 and spin-squared sectors, as noted in \cite{Levi:2014gsa}. In 
addition, terms quadratic in each of the shifts contribute to the NLO 
spin-squared sector. After this shift at LO, one can proceed at NLO to 
eliminate the remaining higher order time derivatives by insertion of EOM, 
where one should use the shifted form of the potential.

\subsubsection{LO quadratic in spin sectors}

In the LO spin1-spin2 and spin-squared sectors, which contain a single Feynman 
diagram each, there is no dependence in the spin gauge, hence there is no change 
in figure 1 of \cite{Levi:2008nh}, nor in figure 1 of \cite{Levi:2014gsa}, 
respectively. The LO spin1-spin2 potential can be found in eq.~(11) of 
\cite{Levi:2008nh}, and reads
\begin{align}
V_{\text{S$_1$S$_2$}}^{\text{LO}}=&-\frac{G}{r^3}\left[\vec{S}_1\cdot\vec{S}_2
-3\vec{S}_1\cdot\vec{n}\vec{S}_2\cdot\vec{n}\right],
\end{align}
where $V=-L$, and the LO spin-squared potential can be found in eq.~(2.15) of 
\cite{Levi:2014gsa}, and reads
\begin{align}
V_{\text{SS}}^{\text{LO}}=&-\frac{C_{{1}{(\text{ES}^2)}}}{2}\frac{Gm_2}
{r^3m_1}\left[S_1^2-3\left(\vec{S}_1\cdot\vec{n}\right)^2\right]
+1\leftrightarrow2.
\end{align}
The Hamiltonians are identical to the potentials, since they are independent
of the velocities. These potentials were first derived in \cite{Barker:1975ae}. 

\subsection{NLO spin-orbit sector} \label{nloso}

Next we compute the NLO spin-orbit sector, for which we essentially have similar Feynman
diagrams as detailed in \cite{Levi:2010zu}, but some changes are found.

First, there is a further kinematic contribution at NLO from eq.~\eqref{frskin}, 
which equals
\be
L_{\text{kin}}=\frac{3}{8}\vec{S}_1\cdot\vec{v}_1\times\vec{a}_1 \,v_1^2
+1\leftrightarrow2.
\ee 
Then, we have 15 Feynman diagrams in total in this sector shown in figure 2.
\begin{figure}[t]
\begin{center}
\includegraphics[width=\textwidth]{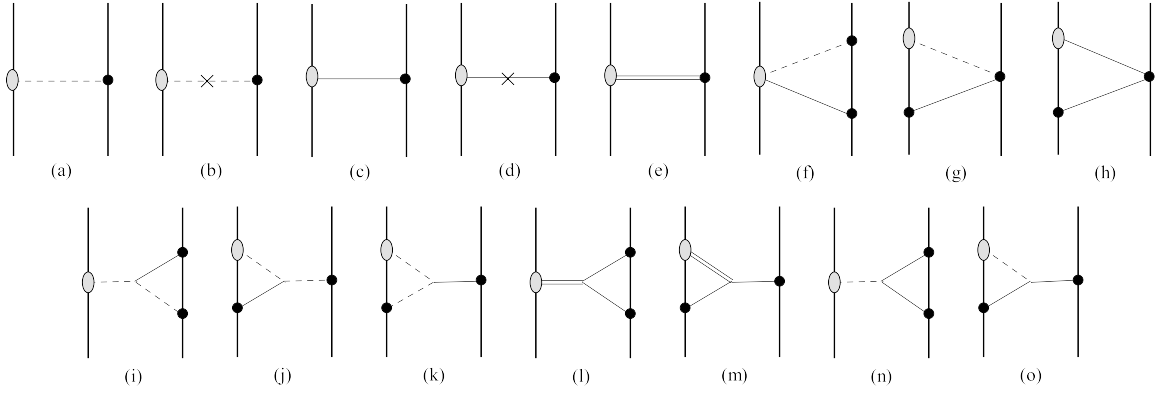}
\caption{NLO spin-orbit Feynman diagrams. These diagrams should be included 
together with their mirror images. Note that the values of diagrams (b), (e), 
(g), and (i)-(o) do not change with respect to \cite{Levi:2010zu}.} 
\end{center}
\label{fig:sonlo}
\end{figure}
Note that the two-scalar spin coupling diagram, which appears as figure 3(b2) 
in \cite{Levi:2010zu}, is absent here, since the corresponding spin coupling 
vanishes in our gauge. The one-loop diagrams, which appear on the bottom row 
of figure 2, do not change their value with respect to the corresponding figure 
4 in \cite{Levi:2010zu}, since they contain only LO spin couplings, which are 
independent of the spin gauge. The values of these diagrams can be found in 
eqs.~(101)-(107) of \cite{Levi:2010zu}. The same is true for diagrams (b), (e), 
and (g), and their values can be found in eqs.~(91), (83), and (94) of 
\cite{Levi:2010zu}, respectively. We should note that in eq.~(91) of 
\cite{Levi:2010zu} terms containing $\vec{a}_2\times\vec{n}$, $\dot{S}_{ij}$ 
were already dropped upon the use of LO EOM. Yet, these contribute at NNLO, 
and moreover all higher order time derivatives should be handled rigorously 
on the same footing in the resulting action \cite{Levi:2011eq,Levi:2014sba}. 
Hence, we rewrite here the complete formal result of diagram (b), reading
\begin{align}
\text{Fig.~2(b)}=&\frac{Gm_2}{r^2}\left[\vec{S}_1\cdot\vec{v}_1\times\vec{v}_2
\,\vec{v}_2\cdot\vec{n}-\vec{S}_1\cdot\vec{v}_2\times\vec{n}\,
\vec{v}_1\cdot\vec{v}_2+3\vec{S}_1\cdot\vec{v}_2\times\vec{n}
\,\vec{v}_1\cdot\vec{n}\,\vec{v}_2\cdot\vec{n}\right]\nn\\
&+\frac{Gm_2}{r}\left[\vec{S}_1\cdot\vec{v}_1\times\vec{a}_2
+\vec{S}_1\cdot\vec{a}_2\times\vec{n}\,\vec{v}_1\cdot\vec{n}
-\dot{\vec{S}}_1\cdot\vec{v}_2\times\vec{n}\,\vec{v}_2\cdot\vec{n}\right]\nn\\
&-Gm_2\dot{\vec{S}}_1\cdot\vec{a}_2\times\vec{n}. 
\end{align}   
The following diagrams change value due to the spin gauge, and they are 
evaluated here as 
\begin{align}
\text{Fig.~2(a)}=&-\frac{Gm_2}{r^2}\left[3\vec{S}_1\cdot\vec{v}_1\times\vec{n}
\,\vec{v}_1\cdot\vec{v}_2+\vec{S}_1\cdot\vec{v}_2\times\vec{n}\,v_2^2
+\vec{S}_1\cdot\vec{v}_1\times\vec{v}_2\,\vec{v}_1\cdot\vec{n}\right]\nn\\
&+4\frac{Gm_2}{r}\left[\vec{S}_1\cdot\vec{a}_1\times\vec{v}_2+
\dot{\vec{S}}_1\cdot\vec{v}_1\times\vec{v}_2\right],\\ 
\text{Fig.~2(c)}=&\left(v_1^2+3v_2^2\right)\frac{Gm_2}{r^2}\vec{S}_1\cdot
\vec{v}_1\times\vec{n} + 2\frac{Gm_2}{r}\vec{S}_1\cdot\vec{v}_1\times
\vec{a}_1,\\
\text{Fig.~2(d)}=&\frac{Gm_2}{r^2}\left[\vec{S}_1\cdot\vec{v}_1\times\vec{n}
\,\vec{v}_1\cdot\vec{v}_2+\vec{S}_1\cdot\vec{v}_1\times\vec{v}_2
\,\vec{v}_1\cdot\vec{n}-3\vec{S}_1\cdot\vec{v}_1\times\vec{n}
\,\vec{v}_1\cdot\vec{n}\,\vec{v}_2\cdot\vec{n}\right]\nn\\
&+\frac{Gm_2}{r}\left[\vec{S}_1\cdot\vec{v}_2\times\vec{a}_1
+\vec{S}_1\cdot\vec{a}_1\times\vec{n}\,\vec{v}_2\cdot\vec{n}
-\dot{\vec{S}}_1\cdot\vec{v}_1\times\vec{v}_2
+\dot{\vec{S}}_1\cdot\vec{v}_1\times\vec{n}\,\vec{v}_2\cdot\vec{n}\right],\\ 
\text{Fig.~2(f)}=&8\frac{G^2m_2^2}{r^3}\vec{S}_1\cdot\vec{v}_2\times\vec{n},\\ 
\text{Fig.~2(h)}=&-2\frac{G^2m_1m_2}{r^3}\vec{S}_1\cdot\vec{v}_1\times\vec{n}. 
\end{align}

\paragraph{The potential.}

Summing all contributions from the kinematic term and Feynman diagrams, we 
obtain the following potential: 
\begin{align} \label{potnloso}
V_{\text{SO}}^{\text{NLO}}=& 
-\frac{G m_2}{r^2} \left[ \vec{S}_1\cdot\vec{v}_1\times\vec{n}\,v_1^2
	 - 2 \vec{S}_1\cdot\vec{v}_1\times\vec{n}\, \vec{v}_1\cdot\vec{v}_2
	 + \vec{S}_1\cdot\vec{v}_1\times\vec{n}\,v_2^2
	  \right.\nl
	 - 3 \vec{S}_1\cdot\vec{v}_1\times\vec{n}\,\vec{v}_1\cdot\vec{n}\,
	 \vec{v}_2\cdot\vec{n}
	 + \vec{S}_1\cdot\vec{v}_2\times\vec{n}\, \vec{v}_1\cdot\vec{v}_2
	 - \vec{S}_1\cdot\vec{v}_2\times\vec{n} \,v_2^2 \nl
	 \left.
	 + 3 \vec{S}_1\cdot\vec{v}_2\times\vec{n}\, \vec{v}_1\cdot\vec{n} \,
	 \vec{v}_2\cdot\vec{n}
	 + \vec{S}_1\cdot\vec{v}_1\times\vec{v}_2 \,\vec{v}_2\cdot\vec{n}\right]\nl
+ \frac{G^2 m_2^2}{2r^3} \left[\vec{S}_1\cdot\vec{v}_1\times\vec{n}
	 -\vec{S}_1\cdot\vec{v}_2\times\vec{n}\right]
+ \frac{3}{8}v_1^2\vec{S}_1\cdot\vec{a}_1\times\vec{v}_1 \nl
+ \frac{G m_2}{r} \left[
		2 \vec{S}_1\cdot\vec{a}_1\times\vec{v}_1
		- 3 \vec{S}_1\cdot\vec{a}_1\times\vec{v}_2
		+ \vec{S}_1\cdot\vec{a}_2\times\vec{v}_1
		- \vec{S}_1\cdot\vec{a}_1\times\vec{n}\, \vec{v}_2\cdot\vec{n}\right.\nl
	 \left.
	  - \vec{S}_1\cdot\vec{a}_2\times\vec{n} \,\vec{v}_1\cdot\vec{n} 
	 - \dot{\vec{S}}_1\cdot\vec{v}_1\times\vec{n}\, \vec{v}_2\cdot\vec{n}
	 + \dot{\vec{S}}_1\cdot\vec{v}_2\times\vec{n}\, \vec{v}_2\cdot\vec{n}
	 -3\dot{\vec{S}}_1\cdot\vec{v}_1\times\vec{v}_2 \right]\nl 
+G m_2 \,\dot{\vec{S}}_1\cdot\vec{a}_2\times\vec{n}
	 + 1 \leftrightarrow 2.
\end{align}
As we noted in the LO spin-orbit sector, we go on to perform a shift of the positions 
according to eq.~\eqref{positionshift}, and get contributions linear in the shift 
from the 1PN order potential, corresponding to the insertion of 1PN order EOM. 
This contribution reads
\begin{align}
\Delta V_{\text{SO}}^{\text{NLO}}\left(\Delta \vec{y}_I\right)=&
  \frac{1}{4} \vec{S}_1\cdot\vec{v}_1\times\vec{a}_1 \,v_1^2  
+ \frac{G m_2}{r} \left[
	   \frac{3}{2} \vec{S}_1\cdot\vec{v}_1\times\vec{a}_1
	 - \frac{7}{4} \vec{S}_1\cdot\vec{v}_1\times\vec{a}_2 \right.\nl \left.
	 - \frac{1}{4} \vec{S}_1\cdot\vec{v}_1\times\vec{n} \,\vec{a}_2\cdot\vec{n}
   \right] 
+ \frac{G m_2}{r^2} \left[
	 \frac{3}{4} \vec{S}_1\cdot\vec{v}_1\times\vec{n} \,v_1^2
	 - 2 \vec{S}_1\cdot\vec{v}_1\times\vec{n}\,\vec{v}_1\cdot\vec{v}_2\right.\nl
	 \left.
	 + \vec{S}_1\cdot\vec{v}_1\times\vec{n}\,v_2^2 
	 + 2 \vec{S}_1\cdot\vec{v}_1\times\vec{v}_2 \,\vec{v}_1\cdot\vec{n} 
	 - \frac{3}{2} \vec{S}_1\cdot\vec{v}_1\times\vec{v}_2\,\vec{v}_2\cdot\vec{n}
	 \right.\nl\left.
	 - \frac{3}{4} \vec{S}_1\cdot\vec{v}_1\times\vec{n}
	 \left(\vec{v}_2\cdot\vec{n}\right)^2\right]
- \frac{G^2 m_2\left(m_1+m_2\right)}{2r^3} \vec{S}_1\cdot\vec{v}_1\times\vec{n}  	 
+ 1 \leftrightarrow 2.
\end{align}
After the shift from LO, we proceed to eliminate the remaining higher order 
time derivatives by insertion of EOM, where we use the shifted form of the 
potential. 
For completeness we present the explicit NLO redefinition of position, which 
removes the higher order time derivatives of position in the potential. It 
reads
\begin{align} \label{nlosopositionshift}
\vec{y}_1 \rightarrow &\vec{y}_1 
+ \frac{1}{8 m_1}\vec{S}_1\times\vec{v}_1 v_1^2
+\frac{G m_2}{m_1 r}\left(\frac{1}{2}\vec{S}_1\times\vec{v}_1
  -3\vec{S}_1\times\vec{v}_2
  -\vec{S}_1\times\vec{n}\,\vec{v}_2\cdot\vec{n}\right)\nn\\
&+\frac{G}{r}\left(\frac{11}{4}\vec{S}_2\times\vec{v}_2
  -\vec{S}_2\times\vec{n}\,\vec{v}_2\cdot\vec{n} 
  +\frac{1}{4}\vec{n}\, \vec{S}_2\cdot\vec{v}_2\times\vec{n}\right),
\end{align}
and similarly for particle 2 with $1\leftrightarrow2$. The contribution 
to the action, which is linear in this shift, removes the acceleration 
terms, and is equivalent to substituting in the EOM of the position.

Finally, we note that at NLO spin-orbit there also appear higher order time derivatives 
of spin, of which a generic rigorous treatment was shown in \cite{Levi:2014sba}, 
see section 5 there. According to this treatment, and considering the relevant 
terms in eq.~\eqref{potnloso}, we realize that whereas the position shift of 
eq.~\eqref{positionshift} was formally a 1PN order shift of the position, here a 2PN 
order redefinition of the spin is required. Terms quadratic in the spin shift 
thus contribute only at the next-to-NNLO (NNNLO) level. Therefore, we can 
consider only the linear in spin shift, which amounts to the insertion of the 
EOM of the spin. Here only the LO Newtonian EOM of the spin contribute, that 
is $\dot{S}_{ij}=0$. 
Again, for completeness we present the explicit redefinition of spin, 
which removes its higher order time derivatives in the potential. 
We use the notation from \cite{Levi:2014sba} for the antisymmetric generator 
of rotation $\omega_1^{ij}$, which transforms the spin variable, and reads
\begin{align}
\omega_1^{ij}=& \frac{G m_2}{r}\left(3v_1^i v_2^j 
+v_1^i n^j \,\vec{v}_2\cdot\vec{n} -v_2^i n^j \,\vec{v}_2\cdot\vec{n}
- (i \leftrightarrow j)\right),   
\end{align}
and similarly for particle 2 with $1\leftrightarrow2$. The contribution 
to the action, linear in this shift, removes the precession terms, and is 
equivalent to substituting in the EOM of the spin.

\paragraph{The Hamiltonian.}

At this stage one can perform a straightforward Legendre transform, see 
e.g.~section 6 in \cite{Levi:2014sba}. Then, we obtain the following 
Hamiltonian:
\begin{align} \label{hamnloso}
H^{\text{NLO}}_{\text{SO}} =& 
\frac{G m_2}{r^2} \left[
	 \frac{5}{8} \vec{S}_1\cdot\frac{\vec{p}_1\times\vec{n}}{m_1} \,
	 \frac{p_1^2}{m_1^2}
	 +3 \vec{S}_1\cdot\frac{\vec{p}_1\times\vec{n}}{m_1} \,
	 \frac{\vec{p}_1 \cdot\vec{n}}{m_1} \,\frac{\vec{p}_2\cdot\vec{n}}{m_2} 
	 - \frac{3}{4} \vec{S}_1\cdot\frac{\vec{p}_1\times\vec{n}}{m_1} 
	 \left(\frac{\vec{p}_2\cdot\vec{n}}{m_2}\right)^2\right.\nl
	 - \vec{S}_1\cdot\frac{\vec{p}_2\times\vec{n}}{m_2}\,
	 \frac{\vec{p}_1\cdot\vec{p}_2}{m_1 m_2}
	 - 3 \vec{S}_1\cdot\frac{\vec{p}_2\times\vec{n}}{m_2}\,
	 \frac{\vec{p}_1\cdot\vec{n}}{m_1} \,\frac{\vec{p}_2\cdot\vec{n}}{m_2}
	 + 2 \vec{S}_1\cdot\frac{\vec{p}_1\times\vec{p}_2}{m_1 m_2} \,
	 \frac{\vec{p}_1\cdot\vec{n}}{m_1}\nl
	 \left.- \frac{5}{2}\vec{S}_1\cdot\frac{\vec{p}_1\times\vec{p}_2}{m_1 m_2} 
	 \,\frac{\vec{p}_2\cdot\vec{n}}{m_2}\right]
+\frac{G^2 m_1m_2}{r^3} \left[
	 \frac{7}{2} \vec{S}_1\cdot\frac{\vec{p}_1\times\vec{n}}{m_1} 
	 - 6\vec{S}_1\cdot\frac{\vec{p}_2\times\vec{n}}{m_2} \right] \nl
+\frac{G^2 m_2^2}{r^3} \left[
	 5 \vec{S}_1\cdot\frac{\vec{p}_1\times\vec{n} }{m_1}
	 - \frac{35}{4} \vec{S}_1\cdot\frac{\vec{p}_2\times\vec{n}}{m_2}\right] 
	 + 1 \leftrightarrow 2.
\end{align}
To verify equivalence with the ADM Hamiltonian result, which was first obtained in 
\cite{Damour:2007nc}, after the corresponding EOM were obtained in 
\cite{Tagoshi:2000zg, Faye:2006gx}, we resolve the difference between the result 
in \cite{Damour:2007nc}, and the result here, using 
canonical transformations, see e.g.~section 7 in \cite{Levi:2014sba}. Using 
the generator of canonical transformations given in eq.~(7.8) there with the 
same notations, we find that the difference is resolved with the coefficients 
of the generator being set to
\be
g_1 =\frac{5}{4}, \quad g_3=\frac{3}{4},
\ee
and the remaining coefficients set to zero, hence indeed the equivalence is 
established.

\subsection{NLO spin1-spin2 sector} \label{nlos1s2}

We go on to handle the NLO spin1-spin2 sector, for which we have similar Feynman 
diagrams as in \cite{Levi:2008nh}, but again some of the values of the diagrams 
are modified. 

There are 6 Feynman diagrams in this sector shown in figure 3.
\begin{figure}[t]
\begin{center}
\includegraphics[width=\textwidth]{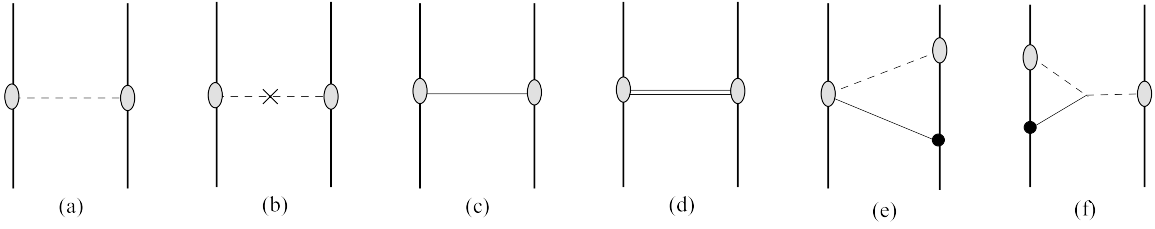}
\caption{NLO spin1-spin2 Feynman diagrams. Diagrams (e) and (f) should be 
included together with their mirror images. Note that the values of diagrams 
(b), (d), and (f) do not change with respect to \cite{Levi:2008nh}.}
\end{center}
\label{fig:s1S2nlo}
\end{figure}
Again, the values of diagrams (b), (d) and (f), in figure 3 do not change with 
respect to \cite{Levi:2008nh}, since they contain only LO spin couplings, 
which are independent of the spin gauge. Their values can be found in eqs.~(25), 
(14), and (32), respectively, of \cite{Levi:2008nh}. Again, in eq.~(25) of 
\cite{Levi:2008nh} terms with time derivatives of spin were already dropped 
upon the use of LO EOM, and hence we rewrite here the complete formal result 
of diagram (b), given by
\begin{align}
\text{Fig.~3(b)}=&-\frac{G}{2r^3}\left[\vec{S}_1\cdot\vec{S}_2\vec{v}_1\cdot
\vec{v}_2-\vec{S}_1\cdot\vec{v}_1\vec{S}_2\cdot\vec{v}_2
-\vec{S}_1\cdot\vec{v}_2\vec{S}_2\cdot\vec{v}_1
-3\vec{S}_1\cdot\vec{S}_2\vec{v}_1\cdot\vec{n}\vec{v}_2\cdot\vec{n}\right.\nn\\
&
+3\vec{S}_1\cdot\vec{v}_1\vec{S}_2\cdot\vec{n}\vec{v}_2\cdot\vec{n}
+3\vec{S}_1\cdot\vec{v}_2\vec{S}_2\cdot\vec{n}\vec{v}_1\cdot\vec{n}
+3\vec{S}_1\cdot\vec{n}\vec{S}_2\cdot\vec{v}_1\vec{v}_2\cdot\vec{n}\nn\\
&\left.
+3\vec{S}_1\cdot\vec{n}\vec{S}_2\cdot\vec{v}_2\vec{v}_1\cdot\vec{n}
+3\vec{S}_1\cdot\vec{n}\vec{S}_2\cdot\vec{n}\vec{v}_1\cdot\vec{v}_2
-15\vec{S}_1\cdot\vec{n}\vec{S}_2\cdot\vec{n}\vec{v}_1\cdot\vec{n}\vec{v}_2
\cdot\vec{n}\right]\nn\\
&-\frac{G}{2r^2}\left[\dot{\vec{S}}_1\cdot\vec{S}_2\vec{v}_2\cdot\vec{n}
-\dot{\vec{S}}_1\cdot\vec{v}_2\vec{S}_2\cdot\vec{n}
-\dot{\vec{S}}_1\cdot\vec{n}\vec{S}_2\cdot\vec{v}_2
+3\dot{\vec{S}}_1\cdot\vec{n}\vec{S}_2\cdot\vec{n}\vec{v}_2\cdot\vec{n}\right]
+1\leftrightarrow2\nn\\
&-\frac{G}{2r}\left[\dot{\vec{S}}_1\cdot\dot{\vec{S}}_2+
\dot{\vec{S}}_1\cdot\vec{n}\dot{\vec{S}}_2\cdot\vec{n}\right]. 
\end{align}
The values of the following diagrams are modified due to the spin gauge, and 
now read
\begin{align}
\text{Fig.~3(a)}=&\frac{G}{2r^3}\left[7\vec{S}_1\cdot\vec{S}_2v_1^2
-7\vec{S}_1\cdot\vec{v}_1\vec{S}_2\cdot\vec{v}_1
-12\vec{S}_1\cdot\vec{S}_2\left(\vec{v}_1\cdot\vec{n}\right)^2
+9\vec{S}_1\cdot\vec{v}_1\vec{S}_2\cdot\vec{n}\vec{v}_1\cdot\vec{n}\right.
\nn\\&\left.
+12\vec{S}_1\cdot\vec{n}\vec{S}_2\cdot\vec{v}_1\vec{v}_1\cdot\vec{n}
-9\vec{S}_1\cdot\vec{n}\vec{S}_2\cdot\vec{n}v_1^2\right]
+1\leftrightarrow2\nn\\
&+2\frac{G}{r^2}\left[\dot{\vec{S}}_1\cdot\vec{S}_2\vec{v}_1\cdot\vec{n}
-\dot{\vec{S}}_1\cdot\vec{n}\vec{S}_2\cdot\vec{v}_1
+\vec{S}_1\cdot\vec{S}_2\vec{a}_1\cdot\vec{n}
-\vec{S}_1\cdot\vec{n}\vec{S}_2\cdot\vec{a}_1\right]+1\leftrightarrow2,\\ 
\text{Fig.~3(c)}=&-8\frac{G}{r^3}\left[\vec{S}_1\cdot\vec{S}_2
\vec{v}_1\cdot\vec{v}_2-\vec{S}_1\cdot\vec{v}_2\vec{S}_2\cdot\vec{v}_1
-\frac{3}{2}\left(
\vec{S}_1\cdot\vec{S}_2\vec{v}_1\cdot\vec{n}\vec{v}_2\cdot\vec{n}-\vec{S}_1
\cdot\vec{v}_2\vec{S}_2\cdot\vec{n}\vec{v}_1\cdot\vec{n}\right.\right.\nn\\
&\left.\left.
-\vec{S}_1\cdot\vec{n}\vec{S}_2\cdot\vec{v}_1\vec{v}_2\cdot\vec{n}
+\vec{S}_1\cdot\vec{n}\vec{S}_2\cdot\vec{n}\vec{v}_1\cdot\vec{v}_2\right)
\right],\\
\text{Fig.~3(e)}=&-4\frac{G^2\left(m_1+m_2\right)}{r^4}\left[\vec{S}_1\cdot
\vec{S}_2-3\vec{S}_1\cdot\vec{n}\vec{S}_2\cdot\vec{n}\right].
\end{align}
We note that in diagrams (c) and (d) double scalar triple products were 
transformed to scalar products according to the appropriate identity, 
e.g.~in eq.~(6.11) of \cite{Levi:2014sba}.

\paragraph{The potential.} \label{potnlos1s2}

Summing all Feynman diagrams, we obtain the following potential: 
\begin{align}
V^{\text{NLO}}_{\text{S$_1$S$_2$}}=& 
-\frac{G}{r^3} \left[
   \frac{7}{2} \vec{S}_1\cdot\vec{S}_2 \,v_1^2	  	
   - \frac{15}{2} \vec{S}_1\cdot\vec{S}_2 \,\vec{v}_1\cdot\vec{v}_2
	 + \frac{7}{2} \vec{S}_1\cdot\vec{S}_2 \,v_2^2
	 - \frac{7}{2} \vec{S}_1\cdot\vec{v}_1 \,\vec{S}_2\cdot\vec{v}_1\right.\nl
	 + \frac{5}{2} \vec{S}_1\cdot\vec{v}_1 \,\vec{S}_2\cdot\vec{v}_2
	 + \frac{9}{2} \vec{S}_1\cdot\vec{v}_2 \, \vec{S}_2\cdot\vec{v}_1
	 - \frac{7}{2} \vec{S}_1\cdot\vec{v}_2 \,\vec{S}_2\cdot\vec{v}_2
	 - 6 \vec{S}_1\cdot\vec{S}_2 \left(\vec{v_1}\cdot\vec{n}\right)^2 \nl
	 + \frac{21}{2} \vec{S}_1\cdot\vec{S}_2\, \vec{v}_1\cdot\vec{n}\, 
	 \vec{v}_2\cdot\vec{n}
	 - 6 \vec{S}_1\cdot\vec{S}_2 \left( \vec{v}_2\cdot\vec{n} \right)^2
	 + \frac{9}{2} \vec{S}_1\cdot\vec{v}_1\, \vec{S}_2\cdot\vec{n}\, 
	 \vec{v}_1\cdot\vec{n} \nl
	 - \frac{9}{2} \vec{S}_1\cdot\vec{v}_1 \,\vec{S}_2\cdot\vec{n}\, 
	 \vec{v}_2\cdot\vec{n}
	 - \frac{15}{2} \vec{S}_1\cdot\vec{v}_2 \,\vec{S}_2\cdot\vec{n}\, 
	 \vec{v}_1\cdot\vec{n} 
	 + 6 \vec{S}_1\cdot\vec{v}_2\,\vec{S}_2\cdot\vec{n}\,\vec{v}_2\cdot\vec{n}\nl 
	 + 6 \vec{S}_1\cdot\vec{n} \,\vec{S}_2\cdot\vec{v}_1\, \vec{v}_1\cdot\vec{n}
	 - \frac{15}{2} \vec{S}_1\cdot\vec{n}\, \vec{S}_2\cdot\vec{v}_1\, 
	 \vec{v}_2\cdot\vec{n} 
	 - \frac{9}{2} \vec{S}_1\cdot\vec{n}\, \vec{S}_2\cdot\vec{v}_2\,
	 \vec{v}_1\cdot\vec{n}  \nl 
	 + \frac{9}{2} \vec{S}_1\cdot\vec{n}\, \vec{S}_2\cdot\vec{v}_2\,
	 \vec{v}_2\cdot\vec{n} 
	 - \frac{9}{2} \vec{S}_1\cdot\vec{n}\, \vec{S}_2\cdot\vec{n}\, v_1^2
	 + \frac{21}{2} \vec{S}_1\cdot\vec{n}\, \vec{S}_2\cdot\vec{n}\, 
	 \vec{v}_1\cdot\vec{v}_2 	 
	 - \frac{9}{2} \vec{S}_1\cdot\vec{n}\, \vec{S}_2\cdot\vec{n}\, v_2^2
	 \nl\left.
	 + \frac{15}{2} \vec{S}_1\cdot\vec{n}\, \vec{S}_2\cdot\vec{n}\, 
	 \vec{v}_1\cdot\vec{n} \, \vec{v}_2\cdot\vec{n} \right]
+ 2\frac{G^2 \left(m_1+m_2\right)}{r^4} \left[ \vec{S}_1\cdot\vec{S}_2
	 -4 \vec{S}_1\cdot\vec{n} \,\vec{S}_2\cdot\vec{n}\right] \nl
- \frac{G}{r^2} \left[ 
   2 \vec{S}_1\cdot\vec{S}_2 \,\vec{a}_1\cdot\vec{n}
	 - 2 \vec{S}_1\cdot\vec{S}_2 \,\vec{a}_2\cdot\vec{n}
   + 2 \vec{S}_1\cdot\vec{a}_2 \,\vec{S}_2\cdot\vec{n}
	 - 2 \vec{S}_1\cdot\vec{n}\,\vec{S}_2\cdot\vec{a}_1  \right. \nl
	 + 2 \dot{\vec{S}}_1\cdot\vec{S}_2 \,\vec{v}_1\cdot\vec{n}
	 - \frac{1}{2} \dot{\vec{S}}_1\cdot\vec{S}_2 \,\vec{v}_2\cdot\vec{n}
	 + \frac{1}{2} \vec{S}_1\cdot\dot{\vec{S}}_2 \,\vec{v}_1\cdot\vec{n} 
	 - 2 \vec{S}_1 \cdot\dot{\vec{S}}_2\,\vec{v}_2\cdot\vec{n} 
	 + \frac{1}{2} \dot{\vec{S}}_1\cdot\vec{v}_2 \,\vec{S}_2\cdot\vec{n} \nl
	 - \frac{1}{2} \vec{S}_1\cdot\vec{n}\, \dot{\vec{S}}_2\cdot\vec{v}_1
	 - 2 \dot{\vec{S}}_1\cdot\vec{n}\, \vec{S}_2\cdot\vec{v}_1 
	 + \frac{1}{2} \dot{\vec{S}}_1\cdot\vec{n}\, \vec{S}_2\cdot\vec{v}_2 
	 - \frac{1}{2} \vec{S}_1\cdot\vec{v}_1\,\dot{\vec{S}}_2\cdot\vec{n}
	 + 2 \vec{S}_1\cdot\vec{v}_2 \,\dot{\vec{S}}_2\cdot\vec{n} \nl\left.
	 - \frac{3}{2} \dot{\vec{S}}_1\cdot\vec{n}\, \vec{S}_2\cdot\vec{n}\, 
	 \vec{v}_2\cdot\vec{n} 
	 + \frac{3}{2} \vec{S}_1\cdot\vec{n}\,\dot{\vec{S}}_2\cdot\vec{n}\,  
	 \vec{v}_1\cdot\vec{n}\right] 
+ \frac{G}{2r} \left[ \dot{\vec{S}}_1\cdot\dot{\vec{S}}_2
	 + \dot{\vec{S}}_1\cdot\vec{n}\, \dot{\vec{S}}_2\cdot\vec{n} \right]. 	 
\end{align}
As noted, we go on to perform a shift of the positions according to 
eq.~\eqref{positionshift}, and get contributions linear in each of the shifts 
from the Newtonian and LO spin-orbit potentials, corresponding to the insertion of 
the EOM from these sectors. This contribution reads 
\begin{align}
\Delta V_{\text{S$_1$S$_2$}}^{\text{NLO}}\left(\Delta \vec{y}_I\right)=&
\frac{G}{r^3} \left[ 
	 \vec{S}_1\cdot\vec{S}_2 \,v_1^2
	 - \frac{3}{2}\vec{S}_1\cdot\vec{S}_2\, \vec{v}_1\cdot\vec{v}_2
	 + \vec{S}_1\cdot\vec{S}_2\, v_2^2
   - \vec{S}_1\cdot\vec{v}_1\, \vec{S}_2\cdot\vec{v}_1 \right.\nl
	 + \frac{3}{2} \vec{S}_1\cdot\vec{v}_2 \,\vec{S}_2\cdot\vec{v}_1 
	 - \vec{S}_1\cdot\vec{v}_2\, \vec{S}_2\cdot\vec{v}_2
	 - \frac{3}{4} \vec{S}_1\cdot\vec{S}_2 \,\vec{v}_1\cdot\vec{n}\,
	 \vec{v}_2\cdot\vec{n} \nl
	 + 3 \vec{S}_1\cdot\vec{v}_1 \,\vec{S}_2\cdot\vec{n}\, 
	 \vec{v}_1\cdot\vec{n}
	 - \frac{9}{4} \vec{S}_1\cdot\vec{v}_2 \,\vec{S}_2\cdot\vec{n}\, 
	 \vec{v}_1\cdot\vec{n} 
	 - \frac{9}{4} \vec{S}_1\cdot\vec{n}\, \vec{S}_2\cdot\vec{v}_1\, 
	 \vec{v}_2\cdot\vec{n}\nl
	 + 3 \vec{S}_1\cdot\vec{n}\,\vec{S}_2\cdot\vec{v}_2\,\vec{v}_2\cdot\vec{n}
	 - 3 \vec{S}_1\cdot\vec{n}\, \vec{S}_2\cdot\vec{n}\,v_1^2
	 + \frac{21}{4} \vec{S}_1\cdot\vec{n}\, \vec{S}_2\cdot\vec{n}\,
	 \vec{v}_1\cdot\vec{v}_2\nl\left.
	 - 3 \vec{S}_1\cdot\vec{n}\, \vec{S}_2\cdot\vec{n}\,v_2^2
	 \right].
\end{align}
After the LO shift, we proceed to eliminate the remaining 
higher order time derivatives, including the time derivatives of spin, which 
appear here too, by insertion of EOM, where one should use the shifted form of 
the potential. 
For completeness we present the explicit NLO redefinition of position, which 
removes its higher order time derivatives in the potential. It reads
\begin{align} \label{nlos1s2positionshift}
\vec{y}_1 \rightarrow &\vec{y}_1 
+2\frac{G}{m_1 r^2}\left(\vec{n}\,\vec{S}_1\cdot\vec{S}_2
-\vec{S}_2\,\vec{S}_1\cdot\vec{n}\right),
\end{align}
and similarly for particle 2 with $1\leftrightarrow2$. The contribution 
to the action, linear in this shift, removes the acceleration terms, and is 
equivalent to substituting in the EOM of the position.

Here again only the LO Newtonian EOM of spin contribute. Again, for completeness 
we present the explicit redefinition of spin, which removes its higher 
order time derivatives, using the antisymmetric generator of rotation 
$\omega_1^{ij}$, which reads
\begin{align}
\omega_1^{ij}=& \frac{G}{r^2}\left(\left(2 S_2^{ik} v_1^j n^k 
-\frac{1}{2}S_2^{ik} v_2^j n^k 
-\frac{3}{2}S_2^{ik} n^j v_2^k
+\frac{3}{2}S_2^{ik} n^j n^k \,\vec{v}_2\cdot\vec{n}\right) 
- (i \leftrightarrow j)-S_2^{ij}\,\vec{v}_2\cdot\vec{n}\right),   
\end{align}
and similarly for particle 2 with $1\leftrightarrow2$. The contribution 
to the action, linear in this shift, removes the precession terms, and is 
equivalent to substituting in the EOM of the spin.

\paragraph{The Hamiltonian.} \label{hamnlos1s2}

Again, at this stage one can perform a straightforward Legendre transform to 
obtain the following Hamiltonian:
\begin{align}
H^{\text{NLO}}_{\text{S$_1$S$_2$}}=&  -\frac{G}{r^3} \left[
   \frac{5}{2} \vec{S}_1\cdot\vec{S}_2\, \frac{p_1^2 }{m_1^2} 
   - 6 \vec{S}_1\cdot\vec{S}_2\, \frac{\vec{p}_1\cdot\vec{p}_2}{m_1 m_2}
   + \frac{5}{2} \vec{S}_1\cdot\vec{S}_2 \,\frac{p_2^2}{m_2^2}
   - \frac{5}{2} \frac{\vec{S}_1\cdot\vec{p}_1}{m_1}\, 
   \frac{\vec{S}_2\cdot\vec{p}_1}{m_1}\right.\nl
   + \frac{5}{2} \frac{\vec{S}_1\cdot\vec{p}_1}{m_1}\, 
   \frac{\vec{S}_2\cdot\vec{p}_2}{m_2}
   + 3 \frac{\vec{S}_1\cdot\vec{p}_2}{m_2}\,
   \frac{\vec{S}_2\cdot\vec{p}_1}{m_1}
   - \frac{5}{2} \frac{\vec{S}_1\cdot\vec{p}_2}{m_2}\, 
   \frac{\vec{S}_2\cdot\vec{p}_2}{m_2}
   - 6 \vec{S}_1\cdot\vec{S}_2 \left(\frac{\vec{p}_1\cdot\vec{n}}
   {m_1}\right)^2\nl
   + \frac{45}{4} \vec{S}_1\cdot\vec{S}_2 \,\frac{\vec{p}_1\cdot\vec{n}}{m_1}
   \,\frac{\vec{p}_2\cdot\vec{n}}{m_2}
   - 6 \vec{S}_1\cdot\vec{S}_2 \left( \frac{\vec{p}_2\cdot\vec{n}}
   {m_2}\right)^2
   + \frac{3}{2} \frac{\vec{S}_1\cdot\vec{p}_1}{m_1} \,
   \vec{S}_2\cdot\vec{n}\, \frac{\vec{p}_1\cdot\vec{n}}{m_1}\nl
   - \frac{9}{2} \frac{\vec{S}_1\cdot\vec{p}_1}{m_1}\,\vec{S}_2\cdot\vec{n}\,
   \frac{\vec{p}_2\cdot\vec{n}}{m_2} 
	 - \frac{21}{4} \frac{\vec{S}_1\cdot\vec{p}_2}{m_2}\,\vec{S}_2\cdot\vec{n}\, 
	 \frac{\vec{p}_1\cdot\vec{n}}{m_1}
	 + 6 \frac{\vec{S}_1\cdot\vec{p}_2}{m_2}\, \vec{S}_2\cdot\vec{n}\, 
	 \frac{\vec{p}_2\cdot\vec{n}}{m_2}\nl
	 + 6 \vec{S}_1\cdot\vec{n}\, \frac{\vec{S}_2\cdot\vec{p}_1}{m_1}\, 
	 \frac{\vec{p}_1\cdot\vec{n}}{m_1} 
	 - \frac{21}{4} \vec{S}_1\cdot\vec{n}\, \frac{\vec{S}_2\cdot\vec{p}_1}{m_1}\,
	 \frac{ \vec{p}_2\cdot\vec{n} }{m_2}
   - \frac{9}{2} \vec{S}_1\cdot\vec{n}\, \frac{\vec{S}_2\cdot\vec{p}_2}{m_2}\, 
   \frac{\vec{p}_1\cdot\vec{n}}{m_1} \nl
   + \frac{3}{2} \vec{S}_1\cdot\vec{n}\, \frac{\vec{S}_2\cdot\vec{p}_2}{m_2} \,
   \frac{\vec{p}_2\cdot\vec{n}}{m_2}
   - \frac{3}{2} \vec{S}_1\cdot\vec{n}\, \vec{S}_2\cdot\vec{n}\, 
   \frac{p_1^2}{m_1^2}
	 + \frac{21}{4} \vec{S}_1\cdot\vec{n}\, \vec{S}_2\cdot\vec{n}\, 
	 \frac{\vec{p}_1\cdot\vec{p}_2}{m_1 m_2}\nl\left.
	 - \frac{3}{2}  \vec{S}_1\cdot\vec{n}\, \vec{S}_2\cdot\vec{n}\, 
	 \frac{p_2^2}{m_2^2}
	 + \frac{15}{2} \vec{S}_1\cdot\vec{n}\, \vec{S}_2\cdot\vec{n}\,
	 \frac{ \vec{p}_1\cdot\vec{n}}{m_1}\, 
	 \frac{\vec{p}_2\cdot\vec{n}}{m_2}\right]\nl
+\frac{G^2 \left(m_1+m_2\right)}{r^4} 
\left[ 7 \vec{S}_1\cdot\vec{S}_2 -13 \vec{S}_1\cdot\vec{n}\, 
\vec{S}_2\cdot\vec{n}\right]. 
\end{align}
Again, to show equivalence with the ADM Hamiltonian in \cite{Steinhoff:2008zr}, 
we use the generator of canonical transformations in eq.~(7.6) of 
\cite{Levi:2014sba} with the same notations. We find that the difference is 
resolved with the coefficients of the generator being set to
\be
g_2 =1, \quad g_3 =1,
\ee
and the remaining coefficients set to zero.

\subsection{NLO spin-squared sector} \label{nloss}

Finally, we compute the NLO spin-squared interaction, which was first 
approached in \cite{Porto:2008jj}. Here we treat it by 
means of the current EFT formulation for spin, and in terms of the NRG fields. 
Since the spin-squared coupling is actually no different 
than a mass quadrupole, it is in fact even more advantageous to use the NRG fields in 
this sector than in the linear in spin sectors, since similarly to the 
non-spinning point-mass sectors higher-loop diagrams are removed, e.g.~as in the 
Einstein-Infeld-Hoffmann 1PN order potential, where a one-loop diagram is absent 
with the use of NRG fields \cite{Kol:2007bc}.

We find that there are 6 Feynman diagrams in this sector shown in figure 4.    
\begin{figure}[t]
\begin{center}
\includegraphics[width=\textwidth]{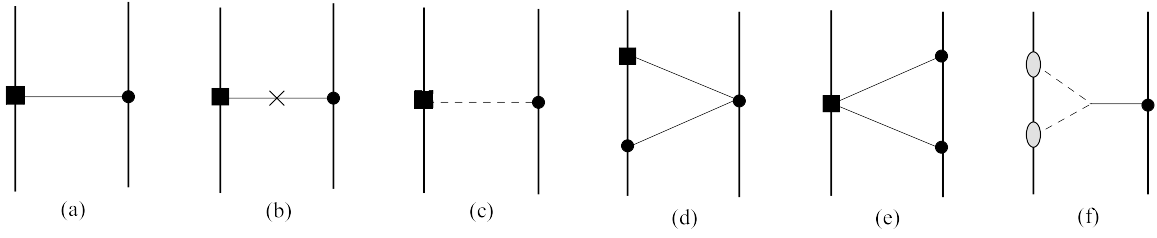}
\caption{NLO spin-squared Feynman diagrams. 
These diagrams should be included together with their mirror images.} 
\end{center}
\label{fig:S2nlo}
\end{figure}
These diagrams are evaluated as follows:
\begin{align}
\text{Fig.~4(a)}=&\frac{C_{1(ES^2)}}{2}\frac{Gm_2}{r^3m_1}\left[
\frac{5}{2}S_1^2v_1^2+\frac{3}{2}S_1^2v_2^2-\left(\vec{S}_1\cdot\vec{v}_1
\right)^2-3S_1^2\left(\vec{v}_1\cdot\vec{n}\right)^2\right.\nn\\
&\left.+3\vec{S}_1\cdot\vec{v}_1\,\vec{S}_1\cdot\vec{n}\,
\vec{v}_1\cdot\vec{n}-\frac{9}{2}\left(\vec{S}_1\cdot\vec{n}\right)^2v_1^2
-\frac{9}{2}\left(\vec{S}_1\cdot\vec{n}\right)^2 v_2^2\right]\nn\\
&+C_{1(ES^2)}\frac{Gm_2}{r^2m_1}\left[\vec{S}_1\cdot\vec{a}_1
\vec{S}_1\cdot\vec{n}
+\dot{\vec{S}}_1\cdot\vec{v}_1\vec{S}_1\cdot\vec{n}
+\dot{\vec{S}}_1\cdot\vec{n}\vec{S}_1\cdot\vec{v}_1
-\dot{S_1^2}\,\vec{v}_1\cdot\vec{n}\right]\nn\\
&+C_{1(ES^2)}\frac{Gm_2}{rm_1}\ddot{S_1^2},\\
\text{Fig.~4(b)}=&-\frac{C_{1(ES^2)}}{4}\frac{Gm_2}{r^3m_1}\left[S_1^2
\vec{v}_1\cdot\vec{v}_2-2\vec{S}_1\cdot\vec{v}_1\vec{S}_1\cdot\vec{v}_2 
-3S_1^2\vec{v}_1\cdot\vec{n}\vec{v}_2\cdot\vec{n}\right.\nn\\
&+6\vec{S}_1\cdot\vec{v}_1\vec{S}_1\cdot\vec{n}\vec{v}_2\cdot\vec{n}
+6\vec{S}_1\cdot\vec{v}_2\vec{S}_1\cdot\vec{n}\vec{v}_1\cdot\vec{n}
+3\left(\vec{S}_1\cdot\vec{n}\right)^2\vec{v}_1\cdot\vec{v}_2\nn\\
&\left.-15\left(\vec{S}_1\cdot\vec{n}\right)^2\vec{v}_1\cdot\vec{n}
\vec{v}_2\cdot\vec{n}\right]+\frac{C_{1(ES^2)}}{4}\frac{Gm_2}{r^2m_1}
\left[2\dot{\vec{S}}_1\cdot\vec{v}_2\vec{S}_1\cdot\vec{n}\right.\nn\\
&\left.
+2\dot{\vec{S}}_1\cdot\vec{n}\vec{S}_1\cdot\vec{v}_2
-6\dot{\vec{S}}_1\cdot\vec{n}\vec{S}_1\cdot\vec{n}\vec{v}_2\cdot\vec{n}
-\dot{S_1^2}\,\vec{v}_2\cdot\vec{n}\right],\\
\text{Fig.~4(c)}=&-2C_{1(ES^2)}\frac{Gm_2}{r^3m_1}\left[S_1^2
-3\left(\vec{S}_1\cdot\vec{n}\right)^2\right]\vec{v}_1\cdot\vec{v}_2\nn\\
&-2C_{1(ES^2)}\frac{Gm_2}{r^2m_1}\left[\dot{\vec{S}}_1\cdot\vec{v}_2
\vec{S}_1\cdot\vec{n}+\dot{\vec{S}}_1\cdot\vec{n}\vec{S}_1\cdot\vec{v}_2
-\dot{S_1^2}\,\vec{v}_2\cdot\vec{n}\right],\\
\text{Fig.~4(d)}=&-C_{1(ES^2)}\frac{G^2m_2}{2r^4}\left[S_1^2-3\left(\vec{S}_1
\cdot\vec{n}\right)^2\right],\\
\text{Fig.~4(e)}=&-2C_{1(ES^2)}\frac{G^2}{r^4}\frac{m_2^2}{m_1}
\left[S_1^2-3\left(\vec{S}_1\cdot\vec{n}\right)^2\right],\\
\text{Fig.~4(f)}=&\frac{G^2m_2}{r^4}\left(\vec{S}_1\cdot\vec{n}\right)^2,
\end{align}
where $\dot{S_1^2}=2\dot{\vec{S}}_1\cdot\vec{S}_1$ is the time derivative of the 
spin length from eq.~\eqref{sldef}, and $\ddot{S_1^2}$ is its second time derivative.
Note that due to the use of NRG fields, we do not have one-loop diagrams with 
the spin-squared coupling since as explained it is actually no different than a 
mass quadrupole. Thus, we do not have here diagrams similar to the two one-loop 
diagrams, which appear in figure 4(a), (b), of \cite{Porto:2008jj}. Therefore, 
the number of diagrams is reduced in this sector too, eliminating in particular 
those, which are more complicated to evaluate.

\paragraph{The potential.} \label{potnloss}

Summing all Feynman diagrams, we obtain the following potential: 
\begin{align}
V_{\text{SS}}^{\text{NLO}}=&  
-\frac{C_{1(ES^2)}}{2} \frac{G m_2}{r^3 m_1} \left[
	 \frac{5}{2} S_1^2 \, v_1^2
	 - \frac{9}{2} S_1^2 \,\vec{v}_1\cdot\vec{v}_2
	 + \frac{3}{2} S_1^2 \,v_2^2
	 - \left(\vec{S}_1\cdot\vec{v}_1\right)^2 
	 + \vec{S}_1\cdot\vec{v}_1 \,\vec{S}_1\cdot\vec{v}_2 \right.\nl
	 -3 S_1^2 \left(\vec{v}_1\cdot\vec{n}\right)^2
	 + \frac{3}{2} S_1^2 \,\vec{v}_1\cdot\vec{n} \,\vec{v}_2 \cdot\vec{n}
	 +3 \vec{S}_1\cdot\vec{v}_1\, \vec{S}_1\cdot\vec{n}\, \vec{v}_1\cdot\vec{n}  
	 -3 \vec{S}_1\cdot\vec{v}_1\, \vec{S}_1\cdot\vec{n}\, \vec{v}_2\cdot\vec{n}
	 \nl
	 -3 \vec{S}_1\cdot\vec{v}_2\, \vec{S}_1\cdot\vec{n}\, \vec{v}_1\cdot\vec{n}  
	 - \frac{9}{2} \left(\vec{S}_1\cdot\vec{n}\right)^2 v_1^2 
	 + \frac{21}{2} \left(\vec{S}_1\cdot\vec{n}\right)^2 \vec{v}_1\cdot\vec{v}_2
	 - \frac{9}{2} \left(\vec{S}_1\cdot\vec{n}\right)^2 v_2^2 
	 \nl \left. 
	 + \frac{15}{2} \left(\vec{S}_1\cdot\vec{n}\right)^2 \vec{v}_1\cdot\vec{n}\, 
	 \vec{v}_2\cdot\vec{n} \right]
+\frac{C_{1(ES^2)}}{2}\frac{G^2 m_2}{r^4} \left[S_1^2
	 -3\left(\vec{S}_1\cdot\vec{n}\right)^2\right] \nl 	 
+2C_{1(ES^2)}\frac{G^2 m_2^2}{r^4m_1}\left[S_1^2
	 -3\left(\vec{S}_1\cdot\vec{n}\right)^2\right] 
-\frac{G^2 m_2}{r^4} \left(\vec{S}_1\cdot\vec{n}\right)^2\nl 
-C_{1(ES^2)}\frac{G m_2}{r^2m_1} \left[
	 \vec{S}_1\cdot\vec{a}_1 \, \vec{S}_1\cdot\vec{n}
	 + \dot{\vec{S}}_1\cdot\vec{v}_1 \, \vec{S}_1\cdot\vec{n} 
	 - \frac{3}{2} \dot{\vec{S}}_1\cdot\vec{v}_2 \, \vec{S}_1\cdot\vec{n}
	 + \dot{\vec{S}}_1\cdot\vec{n} \, \vec{S}_1\cdot\vec{v}_1 \right. \nl\left.
	 - \frac{3}{2} \dot{\vec{S}}_1\cdot\vec{n} \, \vec{S}_1\cdot\vec{v}_2
	 - \frac{3}{2} \dot{\vec{S}}_1\cdot\vec{n} \, \vec{S}_1\cdot\vec{n} \, 
	 \vec{v}_2\cdot\vec{n}
	 - \dot{S_1^2}\, \vec{v}_1\cdot\vec{n}
	 + \frac{7}{4} \dot{S_1^2} \, \vec{v}_2\cdot\vec{n}\right]\nl
-C_{1(ES^2)}\frac{G m_2}{m_1 r} \ddot{S_1^2}
+ 1 \leftrightarrow 2.
\end{align}
We note that we can already take $\dot{S^2}=\ddot{S^2}=0$ since as we noted 
dissipative effects from the absorption of gravitational waves by the compact 
objects, which modify the spin length, are relevant only as of the 5PN order, 
and thus to our approximation the spin length $S^2$ is constant. 
After we make the shift of positions according to eq.~\eqref{positionshift}, we 
get contributions linear in the shifts from the LO spin-orbit potential, equivalent 
to the insertion of EOM from this sector. This contribution reads 
\begin{align}
\Delta V_{\text{SS}}^{\text{NLO}}\left(\Delta \vec{y}_I\right)=&
\frac{G m_2}{r^3 m_1} \left[ 
	 S_1^2 \,v_1^2	
	 - S_1^2 \, \vec{v}_1\cdot\vec{v}_2 
	 - (\vec{S}_1\cdot\vec{v}_1)^2
	 + \vec{S}_1\cdot\vec{v}_1\, \vec{S}_1\cdot\vec{v}_2 
	 + 3 \vec{S}_1\cdot\vec{v}_1 \,\vec{S}_1\cdot\vec{n} \,\vec{v}_1\cdot\vec{n} 
	 \right.\nl\left.
	 - 3 \vec{S}_1\cdot\vec{v}_2 \,\vec{S}_1\cdot\vec{n} \,\vec{v}_1\cdot\vec{n}
	 - 3 (\vec{S}_1\cdot\vec{n})^2\,v_1^2
	 + 3 (\vec{S}_1\cdot\vec{n})^2\,\vec{v}_1\cdot\vec{v}_2\right]\nl
-\frac{1}{2m_1} \left[
	 S_1^2 \,\dot{\vec{a}}_1\cdot\vec{v}_1
	 - \vec{S}_1 \cdot \dot{\vec{a}}_1 \,\vec{S}_1\cdot\vec{v}_1 \right] 
	 + 1 \leftrightarrow 2.
\end{align}
Finally, we note that there is an addition to this sector also from terms 
quadratic in the shift in eq.~\eqref{positionshift}, originating from the Newtonian 
sector. It is given by
\begin{align}
\Delta V_{\text{SS}}^{\text{NLO}}\left((\Delta \vec{y}_I)^2\right)=&
-\frac{Gm_2}{8r^3m_1}\left[
		2 S_1^2\,v_1^2-2(\vec{S}_1\cdot\vec{v}_1)^2
		-3 S_1^2(\vec{v}_1\cdot\vec{n})^2
		+6 \vec{S}_1\cdot\vec{v}_1\,\vec{S}_1\cdot\vec{n}\,\vec{v}_1\cdot\vec{n}
		\right.\nl\left.
		-3(\vec{S}_1\cdot\vec{n})^2v_1^2\right]
-\frac{1}{8m_1}\left[S_1^2\,a_1^2-(\vec{S}_1\cdot\vec{a}_1)^2\right]
+1\leftrightarrow2.
\end{align} 
After this shift, we proceed to eliminate the remaining higher order time 
derivatives, including time derivatives of spin by insertion of the LO 
Newtonian EOM of the spin. 
For completeness we present the explicit NLO redefinition of position, which 
removes its higher order time derivatives. It reads
\begin{align} \label{nlos1s1positionshift}
\vec{y}_1 \rightarrow &\vec{y}_1 
+C_{1(ES^2)}\frac{G m_2}{m_1^2 r^2}\,\vec{S}_1\,\vec{S}_1\cdot\vec{n}
+\frac{G m_2}{8 m_1^2 r^2}\left(\vec{S}_1\,\vec{S}_1\cdot\vec{n}
-\vec{n}\,S_1^2\right)
+\frac{3}{8m_1^2}\left(\vec{S}_1\,\vec{S}_1\cdot\vec{a}_1
-\vec{a}_1\,S_1^2\right),
\end{align}
and similarly for particle 2 with $1\leftrightarrow2$. The contribution 
to the action, linear in this shift, removes the acceleration terms, and is 
equivalent to substituting in the EOM of the position.
For the explicit redefinition of spin we have, using the antisymmetric 
generator of rotation $\omega_1^{ij}$, that
\begin{align} \label{nlos1s1spinredef}
\omega_1^{ij}=& C_{1(ES^2)}\frac{G m_2}{m_1 r^2}
\left(\left(- S_1^{ik} v_1^j n^k - S_1^{ik} n^j v_1^k 
+\frac{3}{2}S_1^{ik} v_2^j n^k 
+\frac{3}{2}S_1^{ik} n^j v_2^k
+\frac{3}{2}S_1^{ik} n^j n^k \,\vec{v}_2\cdot\vec{n}\right)\right.\nn\\ 
&\left.- (i \leftrightarrow j)
+2S_1^{ij}\,\vec{v}_1\cdot\vec{n}
-3S_1^{ij}\,\vec{v}_2\cdot\vec{n}\right)
-\frac{G m_2}{4 m_1 r^2}
\left(S_1^{ik} v_1^j n^k - S_1^{ik} n^j v_1^k - (i \leftrightarrow j)\right)
\nn\\
&-\frac{1}{4 m_1}\left(S_1^{ik} v_1^j a_1^k + S_1^{ik} a_1^j v_1^k
- (i \leftrightarrow j)\right),   
\end{align}
and similarly for particle 2 with $1\leftrightarrow2$. The contribution 
to the action, linear in this shift, removes the precession terms, and is 
equivalent to substituting in the EOM of the spin.
Note that the redefinitions in eqs.~\eqref{nlos1s1positionshift} and 
\eqref{nlos1s1spinredef} contain terms with accelerations. 
These terms are required in order to remove from the potential terms, which 
are quadratic in the accelerations. 

\paragraph{The Hamiltonian.} \label{hamnloss}

Again, one can perform a straightforward Legendre transform to obtain the 
Hamiltonian:
\begin{align}
H^{\text{NLO}}_{\text{SS}} =& 
-\frac{C_{1(ES^2)}}{2} \frac{G m_2}{r^3 m_1} \left[
   \frac{5}{2} S_1^2\, \frac{p_1^2}{m_1^2}
   - \frac{9}{2} S_1^2\, \frac{\vec{p}_1\cdot\vec{p}_2}{m_1 m_2}
   + \frac{3}{2} S_1^2\, \frac{p_2^2}{m_2^2} 
   - \left(\frac{\vec{S}_1\cdot\vec{p}_1}{m_1}\right)^2\right.\nl
   + \frac{\vec{S}_1\cdot\vec{p}_1}{m_1}\,\frac{\vec{S}_1\cdot\vec{p}_2}{m_2}
   - 3 S_1^2 \left(\frac{\vec{p}_1\cdot\vec{n}}{m_1}\right)^2
   + \frac{3}{2} S_1^2 \,\frac{\vec{p}_1\cdot\vec{n}}{m_1}\, \frac{\vec{p}_2\cdot\vec{n}}{m_2}
	 + 3 \frac{\vec{S}_1\cdot\vec{p}_1}{m_1}\,\vec{S}_1\cdot\vec{n}\, 
	 \frac{\vec{p}_1\cdot\vec{n}}{m_1}\nl
	 - 3 \frac{\vec{S}_1\cdot\vec{p}_1}{m_1}\, \vec{S}_1\cdot\vec{n}\,
	 \frac{\vec{p}_2\cdot\vec{n}}{m_2}
	 - 3 \frac{\vec{S}_1\cdot\vec{p}_2}{m_2} \,\vec{S}_1\cdot\vec{n}\,
	 \frac{\vec{p}_1\cdot\vec{n}}{m_1} 
	 - \frac{9}{2} (\vec{S}_1\cdot\vec{n})^2 \frac{p_1^2}{m_1^2}
	 + \frac{21}{2} (\vec{S}_1\cdot\vec{n})^2 
	 \frac{\vec{p}_1\cdot\vec{p}_2}{m_1 m_2} \nl\left.
	 - \frac{9}{2} (\vec{S}_1\cdot\vec{n})^2 \frac{p_2^2}{m_2^2}
	 + \frac{15}{2} (\vec{S}_1\cdot\vec{n})^2 \frac{\vec{p}_1\cdot\vec{n}}{m_1}\, 
	 \frac{\vec{p}_2\cdot\vec{n}}{m_2}\right]
+\frac{G m_2}{r^3 m_1} \left[
	 \frac{5}{4} S_1^2 \,\frac{p_1^2}{m_1^2}
	 - \frac{3}{2} S_1^2 \, \frac{\vec{p}_1\cdot\vec{p}_2}{m_1 m_2} \right.\nl 
	 - \frac{5}{4} \left(\frac{\vec{S}_1\cdot\vec{p}_1}{m_1}\right)^2
	 + \frac{3}{2} \frac{\vec{S}_1\cdot\vec{p}_1}{m_1}\, 
	 \frac{\vec{S}_1\cdot\vec{p}_2}{m_2}
	 - \frac{9}{8} S_1^2 \left(\frac{\vec{p}_1\cdot\vec{n}}{m_1}\right)^2 
	 + \frac{3}{2} S_1^2 \,\frac{\vec{p}_1\cdot\vec{n}}{m_1}\,
	 \frac{\vec{p}_2\cdot\vec{n}}{m_2} \nl
	 + \frac{15}{4} \frac{\vec{S}_1\cdot\vec{p}_1}{m_1} \,\vec{S}_1\cdot\vec{n} \,
	 \frac{\vec{p}_1\cdot\vec{n}}{m_1}
	 - \frac{3}{2} \frac{\vec{S}_1\cdot\vec{p}_1}{m_1} \,\vec{S}_1\cdot\vec{n} \,
	 \frac{\vec{p}_2\cdot\vec{n}}{m_2} 
	 - 3 \frac{\vec{S}_1\cdot\vec{p}_2}{m_2} \,\vec{S}_1\cdot\vec{n} \, 
	 \frac{\vec{p}_1\cdot\vec{n}}{m_1}\nl\left.
	 -\frac{21}{8} (\vec{S}_1\cdot\vec{n})^2 \frac{p_1^2}{m_1^2}
	 +3 (\vec{S}_1\cdot\vec{n})^2 \frac{\vec{p}_1\cdot\vec{p}_2}{m_1 m_2}
	 \right]\nl
+\frac{C_{1(ES^2)}}{2} \frac{G^2 m_2}{r^4} \left[ S_1^2
	 - 3\left( \vec{S}_1\cdot\vec{n} \right)^2 \right] 	 
+C_{1(ES^2)}\frac{G^2 m_2^2}{r^4 m_1} \left[ 2S_1^2
	 - 5 \left( \vec{S}_1\cdot\vec{n}\right)^2 \right] \nl
+\frac{G^2 m_2}{r^4} \left[2S_1^2-3\left(\vec{S}_1\cdot\vec{n}\right)^2\right]  
+\frac{G^2 m_2^2}{r^4 m_1} \left[ S_1^2- \left(\vec{S}_1\cdot\vec{n}\right)^2
\right] + 1 \leftrightarrow 2.
\end{align}
For the NLO spin-squared sector we need to construct the general infinitesimal 
generator of canonical transformations in order to show equivalence with the 
ADM Hamiltonian in \cite{Hergt:2010pa}. Similar to the considerations in 
section 7 of \cite{Levi:2014sba}, we have the following general form for this 
generator:
\begin{align}
g_{\text{SS}}^{\text{NLO}}=&\frac{G m_2}{r^2 m_1}\left[
   S_1^2\left(g_1 \,\frac{\vec{p}_1\cdot\vec{n}}{m_1} 
   +g_2 \,\frac{\vec{p}_2\cdot\vec{n}}{m_2}\right)
	+ \vec{S}_1\cdot\vec{n}\left(g_3\,\frac{\vec{S}_1\cdot\vec{p}_1}{m_1}
	+g_4\,\frac{\vec{S}_1\cdot\vec{p}_2}{m_2}\right)\right.\nl
	\left.+(\vec{S}_1\cdot\vec{n})^2\left(g_5\,\frac{\vec{p}_1\cdot\vec{n}}{m_1} 
	+g_6\,\frac{\vec{p}_2\cdot\vec{n}}{m_2}\right)\right], 
\end{align}
where this generator should be taken with $1\leftrightarrow2$. We find then 
that the difference of Hamiltonians is resolved with the coefficients of the 
generator fixed to
\begin{equation}
g_3 = C_{1(ES^2)},
\end{equation}
and the remaining ones fixed to zero.

\section{Conclusions} \label{theendmyfriend}

In this paper we have presented a formulation for gravitating spinning objects in the effective field theory in the post-Newtonian scheme. We aimed to attain accuracy at the 4PN order for rapidly rotating compact objects, and indeed the formulation holds as it stands to this high PN order. Such 
high PN orders are required for the successful detection of gravitational 
radiation. 

A crucial aspect in our EFT formulation for spinning objects is that we indeed eventually obtain an effective action, where all field modes below the orbital scale are integrated out. This is achieved by introducing several new ingredients, on which this work strongly builds. First, we point out the relevant degrees of freedom, taking special notice of the rotational ones, and most importantly the associated symmetries. Building on these symmetries, we introduce the minimal coupling part of the point particle action in terms of gauge rotational variables, and construct the spin-induced nonminimal couplings, where we obtain the LO couplings to all orders in spin. We then introduce the gauge freedom of rotational variables into the point particle action. Altogether, we construct the point particle effective action following symmetry considerations for spinning objects for the first time. Finally, we fix a canonical gauge for the rotational variables, where the unphysical DOFs are eliminated already from the Feynman rules, and all the orbital field modes are conveniently integrated out. 

The EOM of spin are then directly obtained via a proper 
variation of the action, where they take on a simple form. 
Moreover, the corresponding Hamiltonians are also straightforwardly 
obtained from the potentials derived via this formulation, 
due to the canonical gauge fixing of the rotational variables.
The EFT formulation for spin is implemented here to derive all spin dependent 
potentials up to NLO to quadratic level in spin, i.e.~up to the 3PN order 
for rapidly rotating compact objects. 
In particular, proper NLO spin-squared potential and Hamiltonian are also derived.
For these implementations we use the NRG field decomposition, which is found 
to eliminate higher-loop Feynman diagrams also in the spin dependent sectors, 
and facilitates derivations. 
Therefore, with the additional advantageous usefulness of the obtained Hamiltonians, 
which relate to GW observables, the EFT formulation for spin here is ideal for the 
treatment of higher order spin dependent sectors.

In order to complete the spin dependent conservative sector, i.e.~potentials 
to 4PN order, it remains to apply this EFT formulation for spin at NNLO up to quadratic level in spin, which was initiated in \cite{Levi:2011eq}. Indeed, this was recently obtained in \cite{Levi:2015uxa}, and \cite{Levi:2015ixa}, for the spin-orbit, and spin-squared sectors, respectively, in addition to the LO cubic and quartic in spin sector in \cite{Levi:2014gsa}. Further, one may 
proceed to obtain a formulation of an EFT of radiation for spin for the 
radiative sector along the lines of this work. Implementation on the radiative 
sector covering up to 4PN order would then follow. Finally, it is left for 
future research to reach a better physical understanding of the various gauges of the rotational variables, and of the extra term from minimal coupling, which arises from introducing the spin gauge freedom in the action, and contributes to finite size effects with spin.

\acknowledgments

This work has been done within the Labex ILP (reference ANR-10-LABX-63) part of 
the Idex SUPER, and received financial French state aid managed by the Agence 
Nationale de la Recherche, as part of the programme Investissements d'Avenir 
under the reference ANR-11-IDEX-0004-02.

\bibliographystyle{jhep}
\bibliography{gwbibtex}

\end{document}